\documentclass[11pt]{article}

\usepackage[letterpaper,margin=1in]{geometry}
\usepackage[T1]{fontenc}
\usepackage[utf8]{inputenc}
\usepackage{lmodern}
\usepackage{microtype}
\usepackage{amsmath,amssymb,mathtools}
\usepackage{graphicx}
\usepackage{tikz}
\usepackage{pgfplots}
\usepackage{booktabs}
\usepackage{placeins}
\usepackage{array}
\usepackage{multirow}
\usepackage{enumitem}
\usepackage{algorithm}
\usepackage{algpseudocode}
\usepackage[numbers,sort&compress]{natbib}
\usepackage{xcolor}
\definecolor{amxnavy}{RGB}{9,30,66}
\definecolor{amxblue}{RGB}{84,160,220}
\definecolor{amxgreen}{RGB}{102,194,165}
\definecolor{amxorange}{RGB}{252,141,98}
\definecolor{headroomblue}{RGB}{68,119,170}
\definecolor{headroomgreen}{RGB}{17,119,51}
\definecolor{headroombrown}{RGB}{170,68,34}
\definecolor{headroomyellow}{RGB}{230,185,55}
\usepackage[hidelinks]{hyperref}
\usepackage{titlesec}
\usetikzlibrary{arrows.meta,positioning,fit,calc}
\usepgfplotslibrary{groupplots}
\pgfplotsset{compat=1.17}

\hypersetup{
  pdftitle={BF16 Component-Product Emulation of FP32 and FP64 GEMM on Intel AMX},
  pdfauthor={Cui Bing; Liu Yu},
  pdfsubject={High-performance and mixed-precision matrix multiplication},
  pdfkeywords={Intel AMX, BF16, FP32 and FP64 GEMM, mixed-precision, high-performance computing}
}

\setlist[itemize]{leftmargin=1.6em,itemsep=0.25em,topsep=0.4em}
\titleformat{\section}{\large\bfseries}{\thesection}{0.6em}{}
\titleformat{\subsection}{\normalsize\bfseries}{\thesubsection}{0.6em}{}
\titleformat{\subsubsection}{\normalsize\itshape}{\thesubsubsection}{0.6em}{}

\newcommand{\F}{\mathbb{F}}
\newcommand{\fl}{\operatorname{fl}}
\newcommand{\BFsixteen}{\mathrm{BF16}}
\newcommand{\FBsixtyfour}{\F_{64}^{\mathrm{B}}}
\newcommand{\Fb}{\F_{\BFsixteen}}

\newcommand{\norm}[1]{\left\lVert #1\right\rVert}

\title{\textbf{BF16 Component-Product Emulation\\
of FP32 and FP64 GEMM on Intel AMX}}
\author{
  Cui Bing\textsuperscript{1}, Liu Yu\textsuperscript{1}\thanks{Correspondence to: \href{mailto:liujerry06st@gmail.com}{liujerry06st@gmail.com}}\\[0.35em]
  \textsuperscript{1}\textit{Maginfra Co., Ltd., Jinan, Shandong, China.}
}
\date{}

\begin{document}
\maketitle
\thispagestyle{plain}

\begin{abstract}
Modern CPUs increasingly integrate high-throughput matrix engines optimized for
low-precision AI workloads, while many scientific computing applications still
rely on FP32 and FP64 GEMM to meet their numerical accuracy requirements. This
paper presents a method that combines BF16 matrix products on Intel Advanced
Matrix Extensions (AMX) to emulate real square FP32 and FP64 GEMM. For FP32 GEMM,
each element in both input matrices is decomposed into three BF16 components,
and six selected component products are evaluated to target FP32-level
accuracy comparable to oneMKL SGEMM. For FP64 inputs within the supported
BF16 exponent range, the
method uses a simplified fixed six-slice Ozaki decomposition. Each retained
BF16 product is first produced in FP32, then widened and accumulated in FP64.
Four product-count settings retain 6, 10, 15, or 21 component products,
exposing the accuracy-performance tradeoff relative to
oneMKL DGEMM. The FP32 schedule reuses component data and retains partial
sums in AMX tiles. AMX-FP32 outperforms oneMKL SGEMM at tested orders
$N\ge4096$ on 64 threads and also achieves gains for smaller matrices
at lower thread counts.
AMX-FP64 variants retaining fewer component products can outperform oneMKL DGEMM
at sufficiently large orders; more products improve accuracy at additional cost.
\end{abstract}


\section{Introduction}

The rapid expansion of artificial intelligence has reshaped the design of
modern compute hardware. To meet the arithmetic demand of deep learning,
processor vendors have devoted increasing silicon area and power budget to
low-precision matrix engines. Formats such as FP16, BF16, FP8, and INT8 now
deliver much higher peak throughput than conventional FP32 and, in particular,
FP64 arithmetic on many architectures
\cite{jouppi2017datacenter,choquette2023nvidia,nassif2022sapphire}. This trend
creates a widening imbalance for high-performance computing (HPC). Many
scientific applications, including computational fluid dynamics, climate
simulation, electronic-structure calculations, and quantum chemistry, still
depend on conventional FP32 and FP64 arithmetic for stability, convergence, and
reproducibility.

GEMM is a natural target for exploring whether low-precision matrix hardware
can accelerate computations that require FP32 or FP64 accuracy. In machine learning, GEMM
is the dominant dense linear-algebra primitive; in HPC,
matrix products appear directly in dense solvers and indirectly in sparse
solvers, spectral methods, tensor contractions, electronic-structure methods,
and quantum-chemistry kernels. Modern AI-oriented matrix engines are tile-based:
they exploit locality and parallelism by operating on small matrix blocks, and
they deliver arithmetic density far beyond traditional scalar or vector units.
Low-precision matrix operations often offer substantially higher throughput
than native FP32 or FP64 arithmetic. Exploiting this throughput for computations
requiring higher precision therefore calls for suitable decomposition and
reconstruction methods.

This paper investigates high-precision GEMM emulation on CPUs by combining
selected low-precision matrix products. For FP32 or FP64
GEMM, each element in both input matrices is decomposed into several BF16
components, yielding BF16 component matrices; selected component products are
evaluated by a fast tile engine, and the partial results are reconstructed in a
wider format. The implementation uses Intel Advanced Matrix Extensions (AMX),
which provide BF16 matrix instructions within a general-purpose CPU core.

High-precision emulation is not obtained by issuing additional low-precision
GEMMs alone. Decomposition produces multiple component matrices and, in principle, a
quadratic number of cross products. The most significant products must be
selected according to their numerical contribution, while the accumulation
precision and order must be chosen to meet the target accuracy. On a CPU,
the time spent on conversion, packing, tile configuration,
cache traffic, parallel scheduling, and final reconstruction can be comparable
to that spent on the low-precision matrix operations themselves. A successful design must therefore
couple numerical decomposition with the data movement and blocking structure of
the CPU matrix engine.

This paper presents a BF16-based method for emulating real square FP32 and FP64
GEMM using Intel AMX. The AMX-FP32 path combines a three-component BF16 residual
decomposition with the six-product schedule suggested by Henry et
al.~\cite{henry2019leveraging}. Its numerical accuracy is evaluated alongside
oneMKL SGEMM using a common FP64 reference. The AMX-FP64 path applies a simplified, unscaled Ozaki decomposition to FP64
operands restricted to the BF16 exponent range. It fixes six BF16 slices per
operand, reconstructs each selected component product in FP64, and varies the
number of retained cross-products. Retaining more products typically improves
agreement
with oneMKL DGEMM but reduces the performance headroom available for
acceleration. The implementation combines precomputed component packing with an
AMX-FP32 schedule that reuses component data and retains partial sums in tile
registers. The performance evaluation includes decomposition, packing,
component computation, and reconstruction, and compares the implementations
with oneMKL across matrix sizes and thread counts.

\section{Related Work}

\subsection{Low-Precision Hardware and Mixed-Precision Computing}

Low-precision formats such as BF16\cite{kalamkar2019study} and specialized
matrix engines\cite{markidis2018nvidia} have supported efficient deep-learning
computation and motivated mixed-precision methods for scientific computing.
These methods reserve wider arithmetic for numerically sensitive operations.
In iterative refinement, for
example, a low-precision factorization and solve can be paired with
higher-precision residual evaluation and correction to obtain an accurate linear
system solution\cite{langou2006exploiting}. Carson and Higham established modern
convergence analyses for such schemes and extended them to three-precision
configurations\cite{carson2017new,carson2018accelerating}. At the application
level, mixed-precision methods have reduced computational cost in geostatistical
modeling\cite{abdulah2021accelerating}, reduced data motion in geospatial
modeling\cite{cao2023reducing}, and enabled scale-selective precision in weather
and climate forecasting\cite{chantry2019scale}. These studies demonstrate useful
precision flexibility, but the attainable savings depend on the algorithmic
structure and its error tolerance.

Their objective, however, differs from high-precision GEMM emulation:
assigning formats to different stages of an
algorithm does not recover the information discarded when the input matrices
of a GEMM are rounded to BF16. Reconstructing that GEMM from BF16 component products requires
the discarded low-order information to be represented explicitly in residual
components. This requirement motivates floating-point expansions and precision
emulation.

\subsection{Floating-Point Expansions and Precision Emulation}

Floating-point expansions retain this low-order information explicitly. They
express a value as a sum of floating-point components: a leading component
captures the dominant bits, while residual components retain information
discarded by preceding rounding operations. Classical error-free
transformations, including TwoSum and TwoProduct, separate a
rounded result from its correction term and form the foundation for expansion
arithmetic\cite{dekker1971floating,knuth2014art}. Accurate summation methods
address the complementary problem of combining such terms in the presence of
cancellation and rounding\cite{rump2009accurate}. Multiword arithmetic extends
this idea by representing one number as a sequence of ordinary floating-point
words, although its scalar arithmetic does not automatically benefit from the
throughput of a matrix engine.

For matrix multiplication, an expansion replaces each operand by several
low-precision components and reconstructs the result from selected pairwise
component products. Its accuracy therefore depends on component extraction, the
retained products, and the precision and order of accumulation. RPE, the
low-precision simulator of Higham and Pranesh, and QPyTorch support
format-sensitivity or quantization studies\cite{dawson2017rpe,higham2019simulating,zhang2019qpytorch},
but they do not decompose input matrices into low-precision components and
reconstruct their product with a high-throughput matrix engine. A more direct
line of work maps floating-point expansions onto matrix engines.

\subsection{Accuracy Recovery with Matrix Engines}

Henry et al. showed that three BF16 components per FP32 operand and
six selected BF16 products can recover FP32-oriented matrix-multiplication
accuracy\cite{henry2019leveraging}. Bayraktar et al. recently studied BF16 Tensor
Core emulation of IEEE-754 SGEMM for scientific-computing settings and reported a GPU
implementation intended for library integration, including treatment of
denormal inputs, with improved numerical and performance characteristics
\cite{bayraktar2026exceeding}. Ootomo and Yokota recovered single-precision
accuracy from FP16 Tensor Core operations by combining leading and residual
products with controlled accumulation\cite{ootomo2022recovering}. Fasi et al.
analyzed multiword matrix multiplication on GPU Tensor Cores, identifying how
component products and summation order influence the resulting error
\cite{fasi2023matrix}. Mukunoki et al. developed Tensor
Core-based SGEMM and DGEMM variants that additionally address accuracy and
reproducibility\cite{mukunoki2020dgemm}.

The Ozaki scheme extends this approach based on decomposition and reconstruction
to higher precision by
decomposing operands into scaled low-precision slices, evaluating several
low-precision GEMMs, and accumulating the results in a wider format
\cite{ozaki2012error}. This formulation enabled DGEMM emulation on integer
matrix-multiplication units\cite{ootomo2024dgemm}, but its practical cost depends
strongly on the scaling policy and the number of retained component products.
Abdelfattah et al. analyzed this integer-arithmetic formulation to quantify
that accuracy-cost dependence\cite{abdelfattah2025integeranalysis}. Uchino et al.
developed performance optimizations
for Ozaki-style integer-engine emulation\cite{uchino2025performance} and later
reported an INT8-engine implementation that evaluates both performance and
energy efficiency\cite{uchino2025int8}.

A subsequent line of work seeks more explicit control over the product count.
Ozaki Scheme II uses an integer modular formulation to limit the GEMMs required
for floating-point emulation\cite{ozaki2025schemeii}, and its accuracy behavior
and product-count requirements have been analyzed explicitly
\cite{uchino2026ozakiierror}. Complementing these cost-control approaches,
recent work has established DGEMM-accuracy guarantees with reduced-precision
Tensor Cores\cite{schwarz2026guaranteed}. The same precision-recovery theme has
also been extended to FP8 Tensor Cores with FP64 arithmetic emulation
\cite{mukunoki2026dgemm}, FP8 quantization within Ozaki Scheme II
\cite{uchino2026fp8ozakii}, and reduced intermediate precision requirements in
quantum chemistry\cite{dawson2024reducing}.

Endo et al. optimized native BF16 GEMM on Intel AMX through blocking and
tile-register optimizations\cite{endo2026optimization}. Kouya implemented Ozaki
Scheme II on CPUs using Intel AMX-INT8 and Arm SVE2-i8mm, combining
residue-number representations with Chinese remainder theorem reconstruction
for multiple-precision matrix multiplication\cite{kouya2026ozakiicpu}.

This work combines multiple BF16 component products to emulate GEMM with FP32
and FP64 inputs, rather than optimizing native BF16 GEMM alone or using
integer-based modular reconstruction. Building on the decomposition schemes
of Henry et al. and Ozaki, it focuses on their implementation and evaluation
on Intel AMX.

\section{Background and Problem Setting}

\subsection{Floating-Point Formats}

Table~\ref{tab:formats} summarizes the formats relevant to this work.  We use
$p$ for significand precision including the implicit leading bit and $u=2^{-p}$
for unit roundoff under round-to-nearest.  BF16 retains the eight-bit exponent
field of FP32 but reduces the significand from 24 to 8 bits.  This design gives
BF16 approximately the same normal exponent range as FP32, which is useful for
decomposition: an FP32 value can be repeatedly rounded to BF16 and have the
rounded component subtracted without first changing its exponent range.

\begin{table}[H]
  \centering
  \caption{Floating-point formats used in this work. The precision $p$ includes
  the implicit leading bit.}
  \label{tab:formats}
  \begin{tabular}{lccc}
    \toprule
    Format & Exponent bits & Precision $p$ & Unit roundoff $u$ \\
    \midrule
    BF16 & 8 & 8 & $2^{-8}$ \\
    FP32 & 8 & 24 & $2^{-24}$ \\
    FP64 & 11 & 53 & $2^{-53}$ \\
    \bottomrule
  \end{tabular}
\end{table}

We use the following notation for finite BF16 values and the two IEEE 754 target
formats:
\begin{equation}
  \begin{aligned}
    \Fb &= \{x:x\text{ is a finite BF16 value}\},\\
    \F_{32} &= \{x:x\text{ is a finite IEEE 754 binary32 value}\},\\
    \F_{64} &= \{x:x\text{ is a finite IEEE 754 binary64 value}\}.
  \end{aligned}
  \label{eq:targetformats}
\end{equation}
The corresponding unit roundoffs are denoted by $u_{\BFsixteen}=2^{-8}$,
$u_{32}=2^{-24}$, and $u_{64}=2^{-53}$.

FP64 has a wider exponent range than BF16. We denote the restricted FP64 input
domain used in this work by
\begin{equation}
  \FBsixtyfour =
  \left\{x\in\F_{64}:
  x=0\ \text{or}\ 2^{-126}\leq |x|\leq
  (2-2^{-7})2^{127}\right\}.
  \label{eq:bf16rangefp64}
\end{equation}
Thus, $\FBsixtyfour$ contains FP64 values whose finite nonzero magnitudes lie
within the normal BF16 range; the values retain the 53-bit FP64 significand and
are not rounded to BF16 by this definition. FP64 inputs in this restricted
domain are referred to as \emph{BF16-range FP64} below. This range covers the
operands of many HPC matrix kernels, particularly after normalization or
nondimensionalization. The restriction avoids explicit per-panel scaling and
the associated reconstruction overhead.

This input-range restriction alone does not guarantee representable residual
slices. The unscaled implementation additionally requires every nonzero
extracted component to be exactly representable as a normal BF16 value; zero
components are allowed. Cases violating these conditions are unsupported, as
no scaling or fallback path is implemented.

\subsection{Problem Definition and Performance Model}

This work considers only the square matrix product
\begin{equation}
  C = AB,
  \qquad
  A,B\in\mathcal D_t^{N\times N},\quad
  C\in\F_t^{N\times N},
  \label{eq:gemm}
\end{equation}
where $\F_t=\F_{32}$ and $\mathcal D_t=\F_{32}$ for the AMX-FP32 path, while
$\F_t=\F_{64}$ and $\mathcal D_t=\FBsixtyfour$ for the BF16-range AMX-FP64 path.
Here $N$ is the common row, column, and reduction dimension, and
$\mathcal D_t$ specifies the input domain. The BF16-range restriction applies
only to the FP64 inputs; the reconstructed output uses FP64 and need not lie
within $\FBsixtyfour$. Rectangular and skinny matrix products are outside the
scope of this study.
The numerical claims assume finite operands with normal nonzero values,
finite intermediate FP32 component results and tile accumulations, and finite
target results. The prototype provides no overflow or underflow protection;
exceptional values and boundary cases require a separate fallback extension.

Let $C_{\mathrm{S}}$ and $C_{\mathrm{D}}$ denote the Intel oneMKL SGEMM and
DGEMM results for the corresponding path-specific inputs and test
configuration. Figure legends abbreviate oneMKL as ``MKL.''
AMX-FP32 targets numerical accuracy comparable to SGEMM, without requiring
bitwise agreement. Both results are evaluated against a common FP64 DGEMM
reference computed from exactly widened FP32 inputs; the accuracy metrics
are defined in Section~\ref{sec:inputs-metrics}. For AMX-FP64, no fixed accuracy
threshold is imposed: variants using 6 to 21 component products are evaluated
for their agreement with $C_{\mathrm{D}}$ and performance relative to DGEMM.

If $N_p$ component products are evaluated, a first-order execution-time model is
\begin{equation}
  T_{\mathrm{emu}} = T_{\mathrm{split}} + T_{\mathrm{pack}}
  + N_p T_{\mathrm{AMX}} + T_{\mathrm{recon}} + T_{\mathrm{parallel}}.
  \label{eq:perfmodel}
\end{equation}
Here $T_{\mathrm{split}}$ and $T_{\mathrm{pack}}$ denote decomposition and
packing time, $T_{\mathrm{AMX}}$ is the time for one BF16 component GEMM, and
$T_{\mathrm{recon}}$ and $T_{\mathrm{parallel}}$ account for target-format
reconstruction and parallel overheads, including scheduling and synchronization.
Acceleration requires the AMX throughput advantage to outweigh both the cost
of multiple component products and the non-AMX overheads. This motivates
omitting low-order products, fusing decomposition with packing, and, for
AMX-FP32, reusing tile-resident operands across products.

\subsection{CPU Matrix Engines and Intel AMX}

Recent Xeon processors incorporate Intel AMX, an x86 instruction-set extension
that adds tile registers and matrix instructions to CPU cores
\cite{nassif2022sapphire}. AMX provides dedicated instructions for explicitly
loading and storing blocked matrix data between memory and tile registers,
including \texttt{TILELOADD} and \texttt{TILESTORED}, together with tile
dot-product instructions\cite{intel2025sdm}. This facility provides eight
two-dimensional tile registers. Each tile register has a total capacity of
1~KiB; the implementation can
configure it with at most 16 rows, each having a logical width of at most
64~bytes. For AMX-BF16, the source tiles contain BF16 operands and the
destination tile contains FP32 accumulators.
AMX operates within a general-purpose CPU core and shares its cache hierarchy,
threading environment, and instruction stream. Effective operand reuse therefore
requires coordination between tile-register allocation and CPU cache blocking.

BF16 is selected for its FP32-like exponent range, which supports the
unscaled decomposition used here within the stated domain. The narrower FP16
exponent range can require additional scaling and reconstruction work.

Figure~\ref{fig:amx-execution-model} summarizes the AMX-BF16 execution
primitive at the architectural level. AMX supplies tile-level matrix
multiply-accumulate primitives rather than a complete GEMM kernel. The
implementation configures the tile shapes and organizes panel packing, blocking,
explicit tile movement, operand reuse, and output reconstruction. Packed BF16
panels are loaded into source tiles with \texttt{TILELOADD}. The $B$ panel must
already use the VNNI-packed layout expected by the BF16 dot-product instruction.
\texttt{TDPBF16PS} computes BF16 dot products
and updates an FP32 destination tile. \texttt{TILESTORED} materializes that tile
in memory when subsequent processing or wider-format reconstruction is required.

\begin{figure}[H]
  \centering
  \definecolor{amxnavy}{RGB}{5,24,61}
  \definecolor{amxblue}{RGB}{210,232,252}
  \definecolor{amxgreen}{RGB}{211,239,218}
  \definecolor{amxorange}{RGB}{255,214,158}
  \resizebox{0.995\linewidth}{!}{%
  \begin{tikzpicture}[
    font=\sffamily\scriptsize,
    >=Latex,
    box/.style={
      draw=amxnavy,
      line width=0.85pt,
      rounded corners=4pt,
      align=center,
      text=amxnavy,
      inner xsep=6pt,
      inner ysep=5pt
    },
    tile/.style={
      box,
      minimum width=3.15cm,
      minimum height=0.82cm,
      fill=amxblue
    },
    acc/.style={
      box,
      minimum width=3.15cm,
      minimum height=0.82cm,
      fill=amxorange!55
    },
    mem/.style={
      box,
      minimum width=3.0cm,
      minimum height=0.82cm,
      fill=white
    },
    op/.style={
      box,
      minimum width=3.0cm,
      minimum height=0.9cm,
      fill=amxgreen
    },
    note/.style={
      align=center,
      font=\sffamily\scriptsize
    },
    flow/.style={-{Latex[length=1.55mm,width=1.1mm]}, line width=0.7pt, draw=amxnavy}
  ]
    \node[mem] (memA) {Packed BF16\\operand tile};
    \node[mem, below=0.82cm of memA] (memB) {Packed BF16\\operand tile};

    \node[tile, right=1.25cm of memA] (srcA) {AMX source tile\\BF16};
    \node[tile, right=1.25cm of memB] (srcB) {AMX source tile\\BF16};

    \node[op, right=1.85cm of $(srcA)!0.5!(srcB)$] (tdp) {
      \texttt{tdpbf16ps}\\
      BF16 $\times$ BF16\\
      $\rightarrow$ FP32 accumulate
    };

    \node[acc, below=0.9cm of tdp] (dst) {AMX destination tile\\FP32 accumulator};

    \node[mem, right=1.25cm of tdp] (store) {Stored FP32\\tile result};
    \node[op, below=0.9cm of store] (sw) {Post-AMX\\reconstruction};

    \draw[flow] (memA.east) -- (srcA.west);
    \draw[flow] (memB.east) -- (srcB.west);
    \draw[flow] (srcA.east) -- ([yshift=0.24cm]tdp.west);
    \draw[flow] (srcB.east) -- ([yshift=-0.24cm]tdp.west);
    \draw[flow] (tdp.south) -- (dst.north);
    \draw[flow] (tdp.east) -- (store.west);
    \draw[flow] (store.south) -- (sw.north);

    \node[box, dashed, fit=(srcA)(srcB)(tdp)(dst), inner sep=8pt,
      label={[font=\sffamily\scriptsize,text=amxnavy]above:AMX tile state inside one CPU core}] {};
  \end{tikzpicture}
  }
  \caption{AMX-BF16 execution primitive. \texttt{TILELOADD} explicitly loads
  BF16 source tiles, \texttt{TDPBF16PS} accumulates their dot products into an
  FP32 destination tile, and \texttt{TILESTORED} writes the accumulated tile to
  memory. Higher-precision behavior is obtained by algorithmic decomposition,
  product scheduling, and reconstruction around this primitive.}
  \label{fig:amx-execution-model}
\end{figure}

\section{BF16-Based High-Precision GEMM}
\label{sec:algorithm}

\subsection{General Decomposition Framework}

The notation below builds on classical floating-point splitting and multiword
arithmetic~\cite{dekker1971floating,knuth2014art,fasi2023matrix}, BF16
residual decomposition for FP32 computation~\cite{henry2019leveraging}, and
matrix-splitting and reconstruction formulations used in Ozaki-type matrix
multiplication schemes~\cite{ozaki2012error,ootomo2024dgemm,uchino2025performance,mukunoki2020dgemm}.
We use a single unscaled notation for matrices in $\F_{32}$ and $\FBsixtyfour$.

For a target-format square input matrix
$X\in\F_{32}^{N\times N}\cup(\FBsixtyfour)^{N\times N}$, construct the BF16
expansion
\begin{equation}
  X = \sum_{r=0}^{s_X-1} X_r + R_X,
  \qquad X_r\in\Fb^{N\times N},
  \label{eq:expansion}
\end{equation}
where $s_X$ is the number of BF16 slices, $X_r$ is the $r$th slice, and $R_X$
is the exact remainder of the stored-slice representation. Here $t=32$ for
FP32 inputs and $t=64$ for BF16-range FP64 inputs. We write
$\fl_{\BFsixteen}(\cdot)$ for entrywise rounding to BF16. Slice
indices increase from the most significant contribution ($r=0$) toward
progressively smaller residual contributions. As is customary in multiword
arithmetic, a slice in an algebraic expression denotes the exact real value of
the stored BF16 number. BF16 values are exactly representable in both FP32 and
FP64, so no explicit format-embedding operator is needed in the formulas below.
Equation~\eqref{eq:expansion} specifies the representation: extraction uses
direct residual rounding for FP32 (Section~\ref{sec:fp32}) and a simplified
Ozaki-style projection for FP64 (Section~\ref{sec:fp64-emulation}).

For a selected index set $\mathcal S$ of component pairs, the reconstruction
used here is written compactly as
\begin{equation}
  \widehat C_t
  = \sum_{(i,j)\in\mathcal S}
    \fl_{32}\!\left(A_iB_j\right)
  \qquad t\in\{32,64\}.
  \label{eq:component-reconstruction}
\end{equation}
Here $A_i,B_j\in\Fb^{N\times N}$.
Each inner $\fl_{32}(A_iB_j)$ denotes an AMX BF16-by-BF16 component GEMM:
its inputs are BF16 and its reduction result is retained in FP32. The sum is
evaluated in target format $t$ and in the implementation's executed order. In
particular, for $t=64$, each FP32 component result is exactly widened before
its FP64 addition; there is no FP32 accumulation across component products.

Later residual slices normally contain lower-order information, so component
pairs are selected in increasing index-sum order.  For $s_A$ and $s_B$ slices,
the triangular selection is
\begin{equation}
  \mathcal S_d=
  \left\{(i,j):0\leq i<s_A,\ 0\leq j<s_B,\ i+j<d\right\},
  \qquad N_p(d)=\left|\mathcal S_d\right|.
  \label{eq:product-selection}
\end{equation}
This ordering is a residual-based heuristic rather than a strict entrywise or
matrix-norm monotonicity guarantee: rounding, cancellation, and the entry
distribution can alter individual product magnitudes.  It nevertheless gives
the triangular truncation used here to trade the number of component products
against accuracy~\cite{henry2019leveraging,fasi2023matrix}.

\subsection{AMX-FP32 Emulation}
\label{sec:fp32}

The FP32 path uses a direct residual decomposition into three BF16 components
per input element. With
$A^{(0)}=A$, define
\begin{equation}
  A_r=\fl_{\BFsixteen}(A^{(r)}),\qquad
  A^{(r+1)}=\fl_{32}\!\left(A^{(r)}-A_r\right),
  \quad r=0,1,2,
  \label{eq:fp32split}
\end{equation}
The BF16 slice $A_r$ is exactly representable when read by FP32 arithmetic, so
the subtraction in Eq.~\eqref{eq:fp32split} introduces no conversion error.
For nonzero normal FP32 residuals away from underflow, $A^{(r)}$ and
$A_r$ have the same sign and are within a factor of two in magnitude.
Sterbenz's lemma therefore makes their FP32 subtraction exact
\cite{higham2002accuracy}. Thus, in this normal case,
$\fl_{32}(A^{(r)}-A_r)$ equals the exact difference; the
$\fl_{32}$ notation is retained to state the implemented FP32 operation and to
cover zero, subnormal, and boundary cases.
$R_A=A-(A_0+A_1+A_2)$ is the residual in exact real arithmetic; $B_r$ and
$R_B$ are defined analogously. Thus
\begin{equation}
  A=A_0+A_1+A_2+R_A,\qquad
  B=B_0+B_1+B_2+R_B.
\end{equation}
The three-component split is paired with the six-product schedule of Henry
et al.~\cite{henry2019leveraging,fasi2023matrix}, referred to here as
``Henry-style.'' Setting $d=3$ in Eq.~\eqref{eq:product-selection} retains the
first three diagonals:
\begin{align}
  \widehat C_{32}={}&
    \fl_{32}(A_0B_0)
    +\fl_{32}(A_0B_1)+\fl_{32}(A_1B_0) \nonumber\\
    &\hspace{20mm}
    +\fl_{32}(A_0B_2)+\fl_{32}(A_1B_1)+\fl_{32}(A_2B_0).
  \label{eq:fp32six}
\end{align}
Equation~\eqref{eq:fp32six} specifies the retained products, which are combined
in FP32 using the execution order in Section~\ref{sec:preprocessing-scheduling}.
The corresponding ideal truncation omits
$A_1B_2$, $A_2B_1$, and $A_2B_2$, as well as
terms containing the residuals.  Under round-to-nearest, normal intermediate
values, and exact residual subtraction, repeated BF16 extraction gives a
componentwise residual reduction of at most $u_{\BFsixteen}=2^{-8}$ per step.
The first omitted diagonal is then of order $u_{\BFsixteen}^3=2^{-24}$ relative
to $|A||B|$.
This truncation estimate motivates the six-product choice but is not a
universal forward-error guarantee; accumulation errors and their dependence on
the input are discussed in Section~\ref{sec:error}.
Algorithm~\ref{alg:fp32} summarizes the decomposition and retained component
products; its loop order is illustrative. The implemented accumulation order
and blocking are described in Sections~\ref{sec:preprocessing-scheduling}
and~\ref{sec:reconstruction-parallelization}.

\begin{algorithm}[H]
  \caption{AMX-FP32 GEMM emulation with a triangular BF16 product schedule}
  \label{alg:fp32}
  \begin{algorithmic}[1]
    \Require $A,B\in\F_{32}^{N\times N}$
    \State $(A_0,A_1,A_2)\gets\operatorname{Split}_{\mathrm{FP32}}(A)$ \Comment{three-component BF16 residual split}
    \State $(B_0,B_1,B_2)\gets\operatorname{Split}_{\mathrm{FP32}}(B)$
    \State $C\gets 0$ in FP32
    \For{$i=0,1,2$}
      \For{$j=0,\ldots,2-i$}
        \State $C\gets\fl_{32}\!\left(C+\fl_{32}(A_iB_j)\right)$
        \Comment{AMX BF16 component GEMM with FP32 accumulation}
      \EndFor
    \EndFor
    \State \Return $C$
  \end{algorithmic}
\end{algorithm}

\subsection{AMX-FP64 Emulation}
\label{sec:fp64-emulation}

The AMX-FP64 path uses a simplified Ozaki residual decomposition with six
BF16 slices per input matrix. Unlike a full Ozaki configuration that selects
decomposition depth and scaling from error bounds to meet a prescribed
accuracy target, this method fixes the depth and varies the number of retained
products to evaluate the accuracy-performance tradeoff.

In the standard Ozaki scheme~\cite{ozaki2012error}, a vector or panel residual
is decomposed into scaled low-precision chunks,
\begin{equation}
  X \approx \sum_{r=0}^{s_X-1} 2^{v_X^{(r)}} \widetilde X^{(r)},
  \qquad \widetilde X^{(r)}\in\Fb^{N\times N},
  \label{eq:ozakiscaled}
\end{equation}
where the exponents $v_X^{(r)}$ are chosen from the magnitude of the current
residual and are used again during reconstruction. The Ozaki add-subtract
projection extracts the high-order residual chunk using FP64 additions and
subtractions.

For supported BF16-range FP64 inputs, the present method stores each extracted
chunk directly as BF16 without a separate reconstruction weight.
The extraction uses FP64 round-to-nearest, ties-to-even arithmetic, with no
overflow and no reassociation of the displayed operations.
For $A^{(0)}=A$, if $A^{(r)}=0$, set $A_r=A^{(r+1)}=0$ and set all
remaining slices to zero. Otherwise, for a nonzero residual, the fixed
six-slice Ozaki split is
\begin{align}
  v_A^{(r)} &=
  \left\lceil\log_2\!\left(\max_{i,j}|A_{ij}^{(r)}|\right)\right\rceil,
  \\
  Z_A^{(r)} &=
  \fl_{64}\!\left(
    \fl_{64}\!\left(A^{(r)}+2^{v_A^{(r)}+\rho}\right)
    -2^{v_A^{(r)}+\rho}
  \right), \nonumber\\
  A_r &= \fl_{\BFsixteen}(Z_A^{(r)}),\qquad
  A^{(r+1)}=\fl_{64}\!\left(A^{(r)}-A_r\right),
  \quad r=0,\ldots,5.
  \label{eq:fp64split}
\end{align}
Here $A_r$ is the retained BF16 slice, which is exactly representable when read
by FP64 arithmetic. The ceiling ensures that
$2^{v_A^{(r)}+\rho}\geq 2^\rho\max_{i,j}|A_{ij}^{(r)}|$, providing the
power-of-two upper bound used in the add-subtract extraction.
For a BF16 slice with an 8-bit significand, we use
$\rho=53-8=45$, the significand-width gap between FP64 and BF16. This choice
sets the extraction granularity so that, under the stated representability
conditions, the extracted chunks can be stored directly in BF16.
Extracted components may be zero; every nonzero component is assumed to be
exactly representable as a normal BF16 value. Hence $A_r=Z_A^{(r)}$.
Under these conditions, the extraction and residual update satisfy
$A^{(r)}=A_r+A^{(r+1)}$ exactly, entrywise. The terminal residual $A^{(6)}$
contains the information not represented by the stored slices and is not
decomposed further. The same recurrence is applied to $B^{(0)}=B$.
The representation is unweighted: the scale is carried in each normal BF16
slice's exponent, so reconstruction sums the slices without separate scale
metadata or rescaling. Components violating the representability conditions
require a scaled or fallback extension.

Computing all six-by-six products would require 36 component GEMMs. To reduce
this cost, the evaluated variants retain only the leading triangular part
of the expansion. For six slices per operand,
\begin{equation}
  N_p(3)=6,\quad N_p(4)=10,\quad
  N_p(5)=15,\quad N_p(6)=21.
  \label{eq:fp64productcounts}
\end{equation}
Let $\mathcal S_d=\{(i,j):0\leq i,j<6,\ i+j<d\}$ denote the selected triangular
set. Let $(i_q,j_q)$, $q=0,\ldots,N_p(d)-1$, list its pairs in the executed
order. At the component-product level, the reconstruction is written as
\begin{equation}
  \begin{aligned}
    C_{64}^{(0)} &= 0,\\
    C_{64}^{(q+1)}
      &= \fl_{64}\!\left(
        C_{64}^{(q)}+
        \operatorname{FP64}\!\left(\fl_{32}(A_{i_q}B_{j_q})\right)
      \right),\\
    \widehat C_{64}^{(d)} &= C_{64}^{(N_p(d))},
  \end{aligned}
  \label{eq:fp64reconstruction}
\end{equation}
for $q=0,\ldots,N_p(d)-1$.
The inner $\fl_{32}$ denotes the AMX BF16-by-BF16 component GEMM with an FP32
result. The operator $\operatorname{FP64}(\cdot)$ denotes its exact widening,
and the outer $\fl_{64}$ denotes rounding in the FP64 accumulator addition.
In the blocked implementation, each FP32 partial result is widened to FP64
and accumulated after processing a reduction panel, without FP32 accumulation
across distinct component products. Algorithm~\ref{alg:fp64} summarizes the
method at the component-product level; blocking and reconstruction details
are given in Section~\ref{sec:reconstruction-parallelization}.

\begin{algorithm}[H]
  \caption{Six-slice AMX-FP64 GEMM approximation using AMX-BF16}
  \label{alg:fp64}
  \begin{algorithmic}[1]
    \Require $A,B\in(\FBsixtyfour)^{N\times N}$, subject to the extraction conditions stated above
    \State $(A_0,\ldots,A_5)\gets\operatorname{Split}_{\mathrm{FP64}}(A)$ \Comment{simplified Ozaki six-slice split}
    \State $(B_0,\ldots,B_5)\gets\operatorname{Split}_{\mathrm{FP64}}(B)$
    \State Choose $(d,N_p)\in\{(3,6),(4,10),(5,15),(6,21)\}$
    \State Order $\mathcal S_d$ as $((i_q,j_q))_{q=0}^{N_p-1}$ in the executed order
    \State $C\gets 0$ in FP64
    \For{$q=0,\ldots,N_p-1$}
      \State $(i,j)\gets(i_q,j_q)$
      \State $C\gets\fl_{64}\!\left(C+\operatorname{FP64}(\fl_{32}(A_iB_j))\right)$
      \Comment{AMX FP32 result; store, widen, and add in FP64}
    \EndFor
    \State \Return $C$
\end{algorithmic}
\end{algorithm}

\section{Numerical Accuracy Considerations}
\label{sec:error}

This section discusses how the number of BF16 slices, the selection of component
products, AMX's FP32 accumulation, and the reconstruction format affect
numerical accuracy.
The attained accuracy is evaluated using the oneMKL DGEMM references defined
in Section~\ref{sec:inputs-metrics} for each input-precision path.

\subsection{Decomposition and Omitted Products}

For AMX-FP32, the three stored BF16 components leave residual matrices $R_A$
and $R_B$ after the direct residual split of Section~\ref{sec:fp32}. The six
selected Henry-style products retain the leading component interactions but
omit lower-order cross-products and all terms involving these residuals.
These omissions and FP32 accumulation both contribute to the final error.
The retained AMX-FP32 products are specified in Eq.~\eqref{eq:fp32six}; the
set-based expression below describes the component-product selections of the AMX-FP64 variants.

The fixed six-slice FP64 split leaves terminal residuals
$A^{(6)}$ and $B^{(6)}$ that are not represented by the stored BF16 components.
Under the stated extraction conditions, the residual updates are exact,
and omitting these terminal residuals is one source of deviation from a native
high-precision GEMM. A separate
algorithmic source is product truncation. If only the index set
$\mathcal S_d=\{(i,j):0\leq i<s_A,\ 0\leq j<s_B,\ i+j<d\}$ is evaluated, an
omitted-product term appears:
\begin{equation}
  E_{\mathrm{omit}}=
  \sum_{\substack{0\leq i<s_A,\ 0\leq j<s_B\\(i,j)\notin\mathcal S_d}}
  A_iB_j.
  \label{eq:omit}
\end{equation}
This term is the main algorithmic difference among AMX-FP64-6, AMX-FP64-10, AMX-FP64-15,
and AMX-FP64-21. Adding triangular diagonals reduces the number of omitted
component products, but does not imply full FP64 accuracy for all matrices. Cancellation, input scaling, and
the residuals left by the fixed six-slice split can still dominate the final
error.

\subsection{Accumulation and Reconstruction}

The products in Eq.~\eqref{eq:omit} are real-arithmetic quantities used only to
analyze slice and truncation error. For each retained component product,
AMX-BF16 multiplies BF16 operands and accumulates the result in FP32 destination
tiles. The multiplication of finite BF16 significands is
exactly representable in FP32 before exponent-range effects, while rounding
occurs during FP32 accumulation and during later combination of component
products. This is the same qualitative issue as in standard floating-point dot
products~\cite{higham2002accuracy}; the exact error depends on the blocking,
reduction order, and number of component products combined in one accumulator.

The two precision paths deliberately make different reconstruction choices. The
AMX-FP32 path combines the six Henry-style products in FP32 to reduce
reconstruction overhead. Partial sums remain in the AMX FP32 accumulator
tiles across the six component products, following the operand-reuse order in
Eq.~\eqref{eq:fp32reuseorder}. This schedule reduces data movement and tile
loads but fixes a particular FP32 summation order across component products.
Its numerical accuracy is evaluated against the common FP64 reference. The
AMX-FP64 variants store each AMX component
result as FP32, exactly embed it into FP64, and then accumulate each selected
product in FP64. Reconstructing the AMX-FP64 path directly in FP32 would discard much of the
recovered low-order information. Accuracy depends on four design choices:
the number of BF16 slices, the retained component-product set, reduction
blocking, and reconstruction precision.

\section{AMX Implementation}
\label{sec:implementation}

\subsection{Blocked Kernel Organization and Tile Mapping}

The implementation follows the conventional layered structure of a
high-performance GEMM. The outer output-block sizes are denoted by $B_M$ and
$B_N$, the reduction-panel length by $B_K$, and the AMX micro-tile dimensions
by $T_M$ and $T_N$.  Outer loops
partition the output matrix into cache-resident $B_M\times B_N$ blocks and
split the reduction dimension into panels of length $B_K$.  Within each output
block, the microkernel updates a $T_M\times T_N$ output tile. The default
configuration is
\begin{equation}
  (B_M,B_N,B_K,T_M,T_N)=(256,256,256,32,32).
\end{equation}
For $N=1024$, the block dimensions are reduced to $B_M=B_N=B_K=128$,
while $T_M=T_N=32$ remain unchanged. All other tested matrix orders use the
default configuration.
The same outer structure is used for FP32 and $\FBsixtyfour$ input emulation;
the difference is the number of packed BF16 panels and the reconstruction path.
The choice $T_M=T_N=32$ is determined by the eight-tile register budget of the
selected microkernel mapping. Two tiles hold the two $16\times32$ row halves of
the $A$ operand, two tiles hold two VNNI-packed $B$ operand panels, and the
remaining four tiles hold the four $16\times16$ FP32 accumulator quadrants.
Thus all eight registers are occupied while updating one $32\times32$ output
micro-tile; this is the largest square component-product update supported by
this register allocation without spilling tile-resident operands or
accumulators.

Figure~\ref{fig:tiles} illustrates this mapping for a $32\times32$
component-product update, including the VNNI layout of the $B_j$ panel.
TMM2 and TMM3 hold the $K=32$ data for output columns $0$--$15$ and
$16$--$31$, respectively.
Each logical $32\times16$ BF16 subpanel is packed into 16 rows of 32 BF16
elements. Adjacent reduction entries for one output column form a pair of
BF16 elements occupying 32 bits, so each packed row contains 16 such pairs
(64 bytes). The two source tiles are therefore divided by output-column
group, not by the reduction dimension. The same tile allocation is used by
the AMX-FP32 and AMX-FP64 paths; they differ in the number of retained
component products and in the reconstruction format after tile stores.

\begin{figure}[H]
  \centering
  \definecolor{amxnavy}{RGB}{5,24,61}
  \definecolor{amxblue}{RGB}{210,232,252}
  \definecolor{amxgreen}{RGB}{211,239,218}
  \definecolor{amxorange}{RGB}{255,214,158}
  \resizebox{0.995\linewidth}{!}{%
  \begin{tikzpicture}[y=1.3cm,
    font=\sffamily\fontsize{9}{10.5}\selectfont,
    >=Latex,
	    frame/.style={
	      draw=amxnavy,
	      line width=0.95pt,
	      rounded corners=5pt,
      fill=white,
      align=center,
    },
	    tile/.style={
	      draw=amxnavy,
	      line width=0.65pt,
      fill=amxblue,
      align=center
    },
	    opbox/.style={
	      draw=amxnavy,
	      line width=0.85pt,
	      rounded corners=4pt,
      fill=blue!4,
      align=center,
      inner sep=6pt
    },
	    gridline/.style={draw=amxnavy!80, line width=0.24pt},
	    flow/.style={-{Latex[length=1.55mm,width=1.1mm]}, line width=0.65pt, draw=amxnavy}
  ]
    \draw[frame] (-0.45,-0.08) rectangle (3.55,5.55);
    \node[font=\sffamily\bfseries\fontsize{9}{10.5}\selectfont, text=amxnavy] at (1.55,5.22)
      {BF16 source tiles};

    \node[font=\sffamily\fontsize{9}{10.5}\selectfont, text=amxnavy] at (1.55,4.82)
      {$A_i$ panel: two row halves};
    \node[font=\sffamily\bfseries\fontsize{9}{10.5}\selectfont, text=amxnavy] at (0.12,4.20) {TMM0};
    \draw[tile, fill=amxblue] (0.75,3.80) rectangle (3.25,4.60);
    \foreach \x in {1,...,7}
      \draw[gridline] ({0.75+2.5*\x/8},3.80) -- ({0.75+2.5*\x/8},4.60);
    \foreach \y in {1,...,3}
      \draw[gridline] (0.75,{3.80+0.8*\y/4}) -- (3.25,{3.80+0.8*\y/4});
    \node[font=\fontsize{9}{10.5}\selectfont, text=amxnavy] at (2.00,3.66) {rows 0--15};

    \node[font=\sffamily\bfseries\fontsize{9}{10.5}\selectfont, text=amxnavy] at (0.12,3.10) {TMM1};
    \draw[tile, fill=amxblue] (0.75,2.70) rectangle (3.25,3.50);
    \foreach \x in {1,...,7}
      \draw[gridline] ({0.75+2.5*\x/8},2.70) -- ({0.75+2.5*\x/8},3.50);
    \foreach \y in {1,...,3}
      \draw[gridline] (0.75,{2.70+0.8*\y/4}) -- (3.25,{2.70+0.8*\y/4});
    \node[font=\fontsize{9}{10.5}\selectfont, text=amxnavy] at (2.00,2.56) {rows 16--31};
    \node[font=\fontsize{9}{10.5}\selectfont, text=amxnavy] at (2.00,2.35)
      {each $16\times32$ BF16};

    \draw[amxnavy!45, line width=0.45pt] (-0.15,2.14) -- (3.25,2.14);
    \node[font=\sffamily\fontsize{9}{10.5}\selectfont, text=amxnavy] at (1.55,1.82)
      {$B_j$ panel: VNNI packed};

    \node[font=\sffamily\bfseries\fontsize{9}{10.5}\selectfont, text=amxnavy] at (0.82,1.43) {TMM2};
    \draw[tile, fill=amxgreen] (0.17,0.62) rectangle (1.47,1.28);
    \foreach \x in {1,...,7}
      \draw[gridline] ({0.17+1.3*\x/8},0.62) -- ({0.17+1.3*\x/8},1.28);
    \foreach \y in {1,...,3}
      \draw[gridline] (0.17,{0.62+0.66*\y/4}) -- (1.47,{0.62+0.66*\y/4});
    \node[font=\fontsize{9}{10.5}\selectfont, align=center, text=amxnavy] at (0.82,0.48) {cols 0--15};

    \node[font=\sffamily\bfseries\fontsize{9}{10.5}\selectfont, text=amxnavy] at (2.45,1.43) {TMM3};
    \draw[tile, fill=amxgreen] (1.80,0.62) rectangle (3.10,1.28);
    \foreach \x in {1,...,7}
      \draw[gridline] ({1.80+1.3*\x/8},0.62) -- ({1.80+1.3*\x/8},1.28);
    \foreach \y in {1,...,3}
      \draw[gridline] (1.80,{0.62+0.66*\y/4}) -- (3.10,{0.62+0.66*\y/4});
    \node[font=\fontsize{9}{10.5}\selectfont, align=center, text=amxnavy] at (2.45,0.48) {cols 16--31};
    \node[font=\fontsize{9}{10.5}\selectfont, text=amxnavy] at (1.62,0.18)
      {each $16\times32$ BF16};

    \node[opbox, minimum width=2.75cm, minimum height=2.15cm] (tdp) at (5.75,2.95)
      {{\fontsize{9}{10.5}\selectfont\textbf{AMX-BF16}}\\[1pt]
       \texttt{tdpbf16ps}\\[1pt]
       {BF16 $\times$ BF16 $\rightarrow$ FP32}\\[1pt]
       \textbf{4 quadrant updates}};

    \draw[frame] (8.05,0.8) rectangle (11.55,5.15);
    \node[font=\sffamily\bfseries\fontsize{9}{10.5}\selectfont, text=amxnavy] at (9.80,4.83)
      {$G_\ell$ accumulator block};
    \node[font=\fontsize{9}{10.5}\selectfont, text=amxnavy] at (9.80,4.50) {$32\times32$ FP32};

    \draw[tile, fill=amxorange] (8.55,1.42) rectangle (11.05,3.96);
    \draw[amxnavy, line width=0.85pt] (9.80,1.42) -- (9.80,3.96);
    \draw[amxnavy, line width=0.85pt] (8.55,2.69) -- (11.05,2.69);
    \foreach \x in {1,...,7}
      \draw[gridline] ({8.55+2.5*\x/8},1.42) -- ({8.55+2.5*\x/8},3.96);
    \foreach \y in {1,...,7}
      \draw[gridline] (8.55,{1.42+2.54*\y/8}) -- (11.05,{1.42+2.54*\y/8});

    \node[fill=white, inner sep=1.0pt, font=\fontsize{9}{10.5}\selectfont, align=center] at (9.18,3.33)
      {TMM4};
    \node[fill=white, inner sep=1.0pt, font=\fontsize{9}{10.5}\selectfont, align=center] at (10.42,3.33)
      {TMM5};
    \node[fill=white, inner sep=1.0pt, font=\fontsize{9}{10.5}\selectfont, align=center] at (9.18,2.05)
      {TMM6};
    \node[fill=white, inner sep=1.0pt, font=\fontsize{9}{10.5}\selectfont, align=center] at (10.42,2.05)
      {TMM7};
    \node[font=\fontsize{9}{10.5}\selectfont, text=amxnavy] at (9.80,1.15)
      {four $16\times16$ FP32 tiles};

    \node[opbox, font=\sffamily\fontsize{9}{10.5}\selectfont, minimum width=1.85cm, minimum height=1.35cm] (store) at (13.45,2.95)
      {tile store\\FP32 block\\reconstruction};

    \draw[flow] (3.55,3.76) -- ([yshift=0.36cm]tdp.west);
    \draw[flow] (3.55,2.05) -- ([yshift=-0.36cm]tdp.west);
    \draw[flow] (tdp.east) -- (8.05,2.95);
    \draw[flow] (11.55,2.95) -- (store.west);
  \end{tikzpicture}
  }
  \caption{Representative AMX tile mapping for one $32\times32$ BF16 component
  product. TMM0--TMM1 hold the two row halves of $A_i$.  TMM2--TMM3 hold the
  VNNI-packed $B_j$ data for output columns $0$--$15$ and $16$--$31$,
  respectively; each occupies $16\times32$ BF16 elements, with adjacent
  reduction entries paired within each row. TMM4--TMM7 hold the four
  $16\times16$ FP32 accumulator quadrants. Grid lines are schematic;
  labels give the actual tile dimensions and index ranges.}
  \label{fig:tiles}
\end{figure}

\subsection{Precomputed Decomposition, Packing, and Product Scheduling}
\label{sec:preprocessing-scheduling}

The BF16 components generated by the decompositions in
Section~\ref{sec:algorithm} are packed before the AMX compute loop. Both paths
write each decomposed subblock to a block-contiguous component buffer, allowing
later panel reads to access consecutive packed subblocks rather than strided
regions of the original matrix.

The $A$ and $B$ component buffers differ only in their AMX-facing layout.
Components of $A$ are stored as contiguous subblocks in the standard row-major
panel layout.  Components of $B$ are first converted to the VNNI layout expected
by the BF16 dot-product instruction, with BF16 pairs contiguous along the
reduction dimension, and are then stored as contiguous VNNI-packed subblocks.
During the GEMM loop, the implementation copies the required $B_M\times B_K$
and $B_K\times B_N$ component panels into block-local AMX-facing buffers and
reuses them across the $32\times32$ micro-tile updates inside the block.
Contiguous panel storage promotes spatial locality, while reuse across
micro-tile updates promotes temporal locality within each output block.
Figure~\ref{fig:preprocess} illustrates this layout using the FP32
three-component path.

\begin{figure}[H]
  \centering
  \definecolor{amxnavy}{RGB}{5,24,61}
  \definecolor{amxblue}{RGB}{210,232,252}
  \definecolor{amxgreen}{RGB}{211,239,218}
  \definecolor{amxorange}{RGB}{255,214,158}
  \resizebox{0.995\linewidth}{!}{%
  \begin{tikzpicture}[y=1.25cm,
    font=\sffamily\fontsize{9}{10.5}\selectfont,
    >=Latex,
    frame/.style={
      draw=amxnavy,
      line width=0.95pt,
      rounded corners=5pt,
      fill=white,
      align=center
    },
    tile/.style={
      draw=amxnavy,
      line width=0.65pt,
      align=center
    },
    opbox/.style={
      draw=amxnavy,
      line width=0.85pt,
      rounded corners=4pt,
      fill=blue!4,
      align=center,
      inner sep=6pt
    },
    gridline/.style={draw=amxnavy!80, line width=0.25pt},
    flow/.style={-{Latex[length=1.55mm,width=1.1mm]}, line width=0.65pt, draw=amxnavy}
  ]
    \draw[frame] (0,0.15) rectangle (3.0,5.25);
    \node[font=\sffamily\bfseries\fontsize{9}{10.5}\selectfont, text=amxnavy] at (1.5,4.95)
      {FP32 input panels};
    \node[font=\fontsize{9}{10.5}\selectfont, text=amxnavy] at (1.5,4.66)
      {input subblocks};

    \node[font=\sffamily\bfseries\fontsize{9}{10.5}\selectfont, anchor=west, text=amxnavy] at (0.72,4.42)
      {$A[m_c,k_c]$};
    \draw[tile, fill=amxblue] (0.91,3.05) rectangle (2.09,4.23);
    \foreach \x in {1,...,7}
      \draw[gridline] ({0.91+1.18*\x/8},3.05) -- ({0.91+1.18*\x/8},4.23);
    \foreach \y in {1,...,7}
      \draw[gridline] (0.91,{3.05+1.18*\y/8}) -- (2.09,{3.05+1.18*\y/8});
    \node[font=\fontsize{9}{10.5}\selectfont, text=amxnavy] at (1.5,2.87) {$256\times256$ FP32};

    \node[font=\sffamily\bfseries\fontsize{9}{10.5}\selectfont, anchor=west, text=amxnavy] at (0.72,2.43)
      {$B[k_c,n_c]$};
    \draw[tile, fill=amxgreen] (0.91,1.08) rectangle (2.09,2.26);
    \foreach \x in {1,...,7}
      \draw[gridline] ({0.91+1.18*\x/8},1.08) -- ({0.91+1.18*\x/8},2.26);
    \foreach \y in {1,...,7}
      \draw[gridline] (0.91,{1.08+1.18*\y/8}) -- (2.09,{1.08+1.18*\y/8});
    \node[font=\fontsize{9}{10.5}\selectfont, text=amxnavy] at (1.5,0.90) {$256\times256$ FP32};

    \node[opbox, minimum width=3.05cm, text width=2.95cm, minimum height=2.55cm] (prep) at (5.12,2.72)
      {{\fontsize{9}{10.5}\selectfont\textbf{preprocessing}}\\[2pt]
       BF16 split\\[1pt]
       $x\mapsto x_0,x_1,x_2$\\[5pt]
       row-major $A$ packing\\[2pt]
       VNNI $B$ packing};

    \draw[frame] (7.45,0.15) rectangle (12.45,5.25);
    \node[font=\sffamily\bfseries\fontsize{9}{10.5}\selectfont, text=amxnavy] at (9.95,4.95)
      {BF16 component buffers};

    \node[font=\sffamily\fontsize{9}{10.5}\selectfont, align=center, text=amxnavy] at (9.95,4.50)
      {$A_i$ component buffers\\contiguous row-major subblocks};
    \foreach \idx/\y/\lab in {0/3.70/{$A_0$},1/3.24/{$A_1$},2/2.78/{$A_2$}} {
      \node[font=\sffamily\bfseries\fontsize{9}{10.5}\selectfont, anchor=east, text=amxnavy] at (8.42,{\y+0.18}) {\lab};
      \foreach \s in {0,...,5} {
        \draw[tile, fill=amxblue!45!white]
          ({8.58+0.45*\s},\y) rectangle ({8.93+0.45*\s},{\y+0.35});
      }
      \draw[tile, fill=amxblue]
        ({8.58+0.45*2},\y) rectangle ({8.93+0.45*2},{\y+0.35});
      \node[font=\fontsize{9}{10.5}\selectfont, text=amxnavy] at (11.48,{\y+0.17}) {\dots};
    }

    \draw[amxnavy!45, line width=0.45pt] (7.75,2.58) -- (12.15,2.58);
    \node[font=\sffamily\fontsize{9}{10.5}\selectfont, align=center, text=amxnavy] at (9.95,2.10)
      {$B_i$ component buffers\\contiguous VNNI-packed subblocks};
    \foreach \idx/\y/\lab in {0/1.36/{$B_0$},1/0.94/{$B_1$},2/0.52/{$B_2$}} {
      \node[font=\sffamily\bfseries\fontsize{9}{10.5}\selectfont, anchor=east, text=amxnavy] at (8.42,{\y+0.13}) {\lab};
      \foreach \s in {0,...,4} {
        \draw[tile, fill=amxgreen!45!white]
          ({8.58+0.58*\s},\y) rectangle ({9.08+0.58*\s},{\y+0.26});
        \draw[gridline] ({8.83+0.58*\s},\y) -- ({8.83+0.58*\s},{\y+0.26});
      }
      \draw[tile, fill=amxgreen]
        ({8.58+0.58*2},\y) rectangle ({9.08+0.58*2},{\y+0.26});
      \draw[gridline] ({8.83+0.58*2},\y) -- ({8.83+0.58*2},{\y+0.26});
      \node[font=\fontsize{9}{10.5}\selectfont, text=amxnavy] at (11.80,{\y+0.13}) {\dots};
    }

    \node[opbox, minimum width=2.05cm, minimum height=1.2cm] (amxread) at (14.35,2.72)
      {AMX loop reads\\$256\times256$ panels\\into local buffers};

    \draw[flow] (3.0,3.75) -- ([yshift=0.52cm]prep.west);
    \draw[flow] (3.0,1.75) -- ([yshift=-0.52cm]prep.west);
    \draw[flow] (prep.east) -- (7.45,2.72);
    \draw[flow] (12.45,2.72) -- (amxread.west);
  \end{tikzpicture}
  }
  \caption{Precomputed decomposition and packing used by the prototype, shown
  for the AMX-FP32 three-component path with the default $256\times256$ panels.
  The AMX-FP64 path uses the same packing
  dataflow after Ozaki-style decomposition produces BF16 components.}
  \label{fig:preprocess}
\end{figure}

Under the current $M$-outer, $N$-middle, and $K$-inner traversal, precomputing
the component buffers avoids repeated decomposition and packing of the same
input panels across output blocks. The implementation accepts the additional
storage cost to avoid this repeated work.

The triangular selection in Eq.~\eqref{eq:product-selection} selects component
products according to the sum of their slice indices, whereas the execution
order can be chosen to enable tile reuse.
In the current implementation, this tile-resident reuse applies to AMX-FP32,
whose six selected products share the FP32 accumulator tiles before storage.
AMX-FP64 instead materializes each component product before FP64 accumulation,
as described in Section~\ref{sec:reconstruction-parallelization}.

For the six-product AMX-FP32 path, rather than evaluating the selected products
in diagonal order, the implementation uses the operand-reuse order
\begin{equation}
  A_1B_1 \rightarrow A_1B_0 \rightarrow A_2B_0
  \rightarrow A_0B_0 \rightarrow A_0B_1 \rightarrow A_0B_2 .
  \label{eq:fp32reuseorder}
\end{equation}
where each label denotes the AMX product of the corresponding BF16 components.
Adjacent products in this chain share exactly one operand component.  Thus the
first product loads both an $A$ component panel and a $B$ component panel, while
each later product loads only the operand component that changes. For one output
micro-tile and one 32-element reduction step within a $B_K$ panel, a component panel occupies two AMX
input tiles; the schedule therefore reduces the six-product inner-loop traffic
from 12 component-panel loads to 7, or from 24 tile-load instructions to 14.
This reduces both instruction overhead and movement of BF16 panels into the
tile register file.
The schedule preserves the selected products and their sum in real arithmetic,
but changing the FP32 accumulation order can change rounding errors. The
accuracy of the executed schedule is assessed in Section~\ref{sec:accuracy-results}.
Figure~\ref{fig:reuse-schedule} illustrates the schedule.

\FloatBarrier
\begin{figure}[H]
  \centering
  \definecolor{amxnavy}{RGB}{5,24,61}
  \definecolor{amxblue}{RGB}{210,232,252}
  \definecolor{amxgreen}{RGB}{211,239,218}
  \definecolor{amxorange}{RGB}{255,214,158}
  \definecolor{loadSOne}{RGB}{65,145,225}
  \definecolor{loadSTwo}{RGB}{93,165,235}
  \definecolor{loadSThree}{RGB}{124,187,242}
  \definecolor{loadSFour}{RGB}{158,208,249}
  \definecolor{loadSFive}{RGB}{193,228,253}
  \definecolor{loadSSix}{RGB}{226,243,255}
  \definecolor{keepSOne}{RGB}{65,164,95}
  \definecolor{keepSTwo}{RGB}{93,187,120}
  \definecolor{keepSThree}{RGB}{125,207,146}
  \definecolor{keepSFour}{RGB}{160,226,175}
  \definecolor{keepSFive}{RGB}{196,240,205}
  \definecolor{keepSSix}{RGB}{228,250,232}
  \definecolor{arrowblue}{RGB}{0,112,192}
  \resizebox{\linewidth}{!}{%
  \begin{tikzpicture}[
    font=\sffamily\scriptsize,
    >=Latex,
    frame/.style={
      draw=amxnavy,
      line width=0.95pt,
      rounded corners=6pt,
      fill=white,
      align=center
    },
    stage/.style={
      draw=amxnavy,
      line width=0.9pt,
      rounded corners=4pt,
      fill=white,
      align=center,
      minimum width=1.85cm,
      minimum height=0.78cm
    },
    action/.style={
      draw=amxnavy!65,
      line width=0.65pt,
      rounded corners=3pt,
      align=center,
      minimum width=1.55cm,
      minimum height=0.46cm,
      font=\scriptsize
    },
    flow/.style={-{Latex[length=1.55mm,width=1.1mm]}, line width=0.7pt, draw=amxnavy},
    reuse/.style={-{Latex[length=1.5mm,width=1.1mm]}, line width=0.6pt, draw=teal!70!black},
    acc/.style={
      draw=amxnavy,
      line width=0.75pt,
      rounded corners=4pt,
      fill=amxorange!32,
      align=center
    }
  ]
    \draw[frame] (0,0) rectangle (15.35,4.78);
    \node[font=\sffamily\bfseries\scriptsize, text=amxnavy] at (7.675,4.47)
      {FP32 six-product operand-reuse schedule};
    \node[font=\sffamily\scriptsize, text=amxnavy] at (7.675,4.19)
      {AMX micro-tile schedule: input panels are updated one at a time while FP32 accumulators remain live};

    \fill[amxblue!12] (0.95,2.31) rectangle (14.65,2.78);
    \fill[amxgreen!12] (0.95,1.82) rectangle (14.65,2.29);
    \foreach \x/\name/\prodtext/\loadtext in {
      1.25/p0/{$A_1B_1$}/{step 1},
      3.75/p1/{$A_1B_0$}/{step 2},
      6.25/p2/{$A_2B_0$}/{step 3},
      8.75/p3/{$A_0B_0$}/{step 4},
      11.25/p4/{$A_0B_1$}/{step 5},
      13.75/p5/{$A_0B_2$}/{step 6}
    } {
      \node[stage] (\name) at (\x,3.46)
        {{\boldmath\scriptsize \prodtext}\\[-1pt]{\scriptsize \loadtext}};
      \draw[-{Latex[length=1.35mm,width=1.0mm]}, draw=amxnavy!45, line width=0.65pt, dashed]
        (\x,3.04) -- (\x,1.36);
    }

    \draw[flow] (p0.east) -- (p1.west);
    \draw[flow] (p1.east) -- (p2.west);
    \draw[flow] (p2.east) -- (p3.west);
    \draw[flow] (p3.east) -- (p4.west);
    \draw[flow] (p4.east) -- (p5.west);

    \node[action, fill=loadSOne] at (1.25,2.55) {load $A_1$};
    \node[action, fill=loadSOne] at (1.25,2.06) {load $B_1$};

    \node[action, fill=keepSTwo] at (3.75,2.55) {keep $A_1$};
    \node[action, fill=loadSTwo] at (3.75,2.06) {load $B_0$};

    \node[action, fill=loadSThree] at (6.25,2.55) {load $A_2$};
    \node[action, fill=keepSThree] at (6.25,2.06) {keep $B_0$};

    \node[action, fill=loadSFour] at (8.75,2.55) {load $A_0$};
    \node[action, fill=keepSFour] at (8.75,2.06) {keep $B_0$};

    \node[action, fill=keepSFive] at (11.25,2.55) {keep $A_0$};
    \node[action, fill=loadSFive] at (11.25,2.06) {load $B_1$};

    \node[action, fill=keepSSix] at (13.75,2.55) {keep $A_0$};
    \node[action, fill=loadSSix] at (13.75,2.06) {load $B_2$};

    \draw[-{Latex[length=1.55mm,width=1.15mm]}, line width=0.85pt, draw=arrowblue]
      (1.15,0.62) -- (14.20,0.62);
    \node[acc, minimum width=12.95cm, minimum height=0.52cm] (accband) at (7.675,1.02)
      {TMM accumulator group: FP32 $C$ tiles remain resident};
    \node[font=\sffamily\scriptsize, text=amxnavy, align=center] at (7.675,0.28)
      {Two component-panel loads for the first product; one changed component-panel load for each later product.};
  \end{tikzpicture}
  }
  \caption{Operand-reuse scheduling for the AMX-FP32 six-product path.  The AMX-FP32
  accumulator tiles remain resident across the chain; only the changed input
  component panel is reloaded after the first product.}
  \label{fig:reuse-schedule}
\end{figure}

\subsection{Target-Format Reconstruction and Parallelization}
\label{sec:reconstruction-parallelization}

For the AMX-FP32 implementation, reconstruction is fused with the blocked compute
loop.  For each $B_M\times B_N$ output block, the implementation allocates a
private FP32 block accumulator.  Within each $B_K$ panel, the implementation
zeros the four FP32 AMX accumulator tiles for a $32\times32$ micro-tile,
evaluates the six scheduled component products across the $B_K$ reduction
block, and stores the four AMX accumulator tiles to a temporary
$32\times32$ FP32 buffer.  This temporary micro-tile is added into the
thread-private $B_M\times B_N$ FP32 block accumulator.  After all $B_K$ panels
have been processed, the completed block is written to the output matrix.

The AMX-FP64 variants use the same blocked traversal and output-block
parallelization. Within each reduction panel, the four FP32 tiles produced for
each selected component product are stored as a $32\times32$ micro-tile,
converted to FP64, and added to the thread-private FP64 block accumulator.
Combining different component products in the AMX FP32 tiles before conversion
would perform their summation in FP32, contrary to the intended FP64
reconstruction. The additional stores, conversions, and FP64 additions reduce
the performance headroom of the AMX-FP64 variants relative to AMX-FP32.
Figure~\ref{fig:reconstruction-paths} compares the two reconstruction paths.

\begin{figure}[!htbp]
  \centering
  \definecolor{amxnavy}{RGB}{5,24,61}
  \definecolor{amxblue}{RGB}{210,232,252}
  \definecolor{amxgreen}{RGB}{211,239,218}
  \definecolor{amxorange}{RGB}{255,214,158}
  \definecolor{amxpurple}{RGB}{226,218,250}
  \definecolor{amxteal}{RGB}{207,238,238}
  \definecolor{amxgray}{RGB}{245,247,250}
  \resizebox{0.995\linewidth}{!}{%
  \begin{tikzpicture}[
    font=\sffamily\fontsize{9}{11}\selectfont,
    frame/.style={draw=amxnavy,line width=0.85pt,rounded corners=5pt,
      fill=white,align=center},
    block/.style={draw=amxnavy!80,line width=0.7pt,rounded corners=3pt,
      fill=white,align=center,inner sep=3pt,minimum height=0.72cm},
    arrow/.style={-{Latex[length=1.7mm,width=1.2mm]},line width=0.7pt,draw=amxnavy}
  ]
    \draw[frame,fill=amxgray] (0,8.45) rectangle (13.4,9.90);
    \node[font=\sffamily\fontsize{9}{11}\selectfont\bfseries]
      at (6.7,9.55) {Shared OpenMP and blocking structure};
    \node[align=center] at (6.7,8.98)
      {Independent $B_M\times B_N$ output blocks; sequential $B_K$ panels;\\
       the same AMX micro-kernel shape};

    \draw[frame] (0,5.05) rectangle (13.4,7.95);
    \node[font=\sffamily\fontsize{10}{12}\selectfont\bfseries,text=amxnavy]
      at (6.7,7.56) {FP32 reconstruction};
    \node[block,fill=amxblue,text width=2.1cm] (products) at (1.55,6.52)
      {Six selected\\BF16 products};
    \node[block,fill=amxorange,text width=2.65cm] (resident) at (5.0,6.52)
      {AMX FP32 tiles\\retain partial sums};
    \node[block,fill=amxgreen,text width=1.65cm] (store32) at (8.4,6.52)
      {Store\\four tiles};
    \node[block,fill=amxpurple,text width=2.05cm] (acc32) at (11.65,6.52)
      {FP32 block\\accumulator};
    \draw[arrow] (products) -- (resident);
    \draw[arrow] (resident) -- (store32);
    \draw[arrow] (store32) -- (acc32);
    \node[text=amxnavy] at (6.7,5.48)
      {One store stage per micro-tile per reduction panel.};

    \draw[frame] (0,0) rectangle (13.4,4.55);
    \node[font=\sffamily\fontsize{10}{12}\selectfont\bfseries,text=amxnavy]
      at (6.7,4.25) {FP64 reconstruction};
    \node[text=amxnavy] at (6.7,3.77)
      {Materialize each component product before FP64 accumulation};
    \foreach \idx/\y/\lab in {one/3.0/$Q_1$,two/2.08/$Q_2$,last/0.94/$Q_{N_p(d)}$} {
      \node[block,fill=amxblue,text width=1.05cm] (q\idx) at (0.9,\y) {\lab};
      \node[block,fill=amxorange,text width=1.9cm] (t\idx) at (3.2,\y)
        {AMX FP32\\tiles};
      \node[block,fill=amxgreen,text width=1.15cm] (s\idx) at (5.75,\y) {Store};
      \node[block,fill=amxteal,text width=2.5cm] (c\idx) at (8.3,\y)
        {Convert\\FP32 to FP64};
      \draw[arrow] (q\idx) -- (t\idx);
      \draw[arrow] (t\idx) -- (s\idx);
      \draw[arrow] (s\idx) -- (c\idx);
    }
    \foreach \doty in {1.40,1.51,1.62}
      \fill[amxnavy] (0.9,\doty) circle[radius=0.45pt];
    \node[block,fill=amxpurple,text width=2.05cm,minimum height=2.85cm]
      (acc64) at (11.65,1.97) {FP64 block\\accumulator};
    \foreach \idx in {one,two,last}
      \draw[arrow] (c\idx.east) -- (acc64.west |- c\idx.east);
    \node[text=amxnavy] at (6.7,0.25)
      {For every selected product: store, convert, and add to the FP64 accumulator.};
  \end{tikzpicture}}
  \caption{Different reconstruction paths under the same block-level
  parallelization.  The AMX-FP32 path keeps the AMX FP32 accumulator tiles live across the
  six scheduled component products, whereas the AMX-FP64 variants store each
  selected AMX-BF16 product, convert it to FP64, and accumulate outside AMX.
  $Q_1,\ldots,Q_{N_p(d)}$ denote the selected component products within a
  reduction panel.}
  \label{fig:reconstruction-paths}
\end{figure}

The current implementation parallelizes three stages with OpenMP.  The $A$ decomposition is
parallelized over $(B_M,B_K)$ blocks with static scheduling, while the
$B$ decomposition and VNNI packing are parallelized over $(B_K,B_N)$ blocks.
The compute stage assigns independent $B_M\times B_N$ output blocks to threads
using a collapsed static loop over the outer $M$ and $N$ block dimensions, so no
two threads update the same output block.  This parallel structure is shared by
the AMX-FP32 and AMX-FP64 paths.  Inside each output block, the current implementation
traverses $B_K$ panels sequentially and copies the precomputed $A$ components
and VNNI-packed $B$ components into block-local buffers before issuing the AMX
micro-kernel.  The AMX tile configuration is loaded before the timed compute
region, and tile state is released after the AMX kernel completes.

\FloatBarrier
\section{Experimental Methodology}
\label{sec:methodology}

\subsection{Hardware and Experimental Configuration}

Table~\ref{tab:platform} summarizes the experimental platform and build
configuration. Unless otherwise noted, the SGEMM and DGEMM baselines use Intel
oneMKL from the same oneAPI environment as the prototype. OpenMP parallel runs
use the Intel OpenMP runtime with \texttt{KMP\_AFFINITY=compact} for compact
thread placement. Frequency policy is kept fixed within each measurement campaign.

\begin{table}[H]
  \centering
  \caption{Experimental platform.}
  \label{tab:platform}
  \begin{tabular}{ll}
    \toprule
    Item & Configuration \\
    \midrule
    Processor & Intel Xeon Platinum 8462Y+ \\
    Sockets & 2 \\
    Operating system & Rocky Linux 9.3 \\
    Compiler & Intel oneAPI 2026.1 \\
    Compiler flags & \texttt{-O3 -march=native -mkl -qopenmp} \\
    BLAS library & Intel oneMKL 2026.1 \\
    Thread binding & compact \\
    \bottomrule
  \end{tabular}
\end{table}

Full-core measurements use 64 OpenMP threads across the two sockets, with
hyper-threading disabled and one thread per physical core. Thread-scaling
experiments vary the number of OpenMP threads as indicated in the corresponding
results. For each configuration, each method performs 10 untimed warm-up runs
followed by 10 timed runs using the same input matrices. SGEMM and DGEMM each
use their respective routine throughout, while every AMX-FP32 and AMX-FP64 run
includes decomposition, packing, component products, and reconstruction. The
reported arithmetic-mean runtime is obtained by dividing the total wall-clock
time of the 10 timed runs by 10.

\subsection{Baselines and Variants}

The principal baselines are the Intel oneMKL SGEMM and DGEMM routines for the
AMX-FP32 and AMX-FP64 paths, respectively.  Table~\ref{tab:variants} summarizes the
emulation variants evaluated in this study. Within each comparison, the methods
use the same input and output matrix layouts, thread count, and timing protocol.

\begin{table}[H]
  \centering
  \caption{Evaluated algorithm variants.}
  \label{tab:variants}
  \resizebox{\textwidth}{!}{%
  \begin{tabular}{lcccc}
    \toprule
    Variant & Performance baseline & Slices & BF16 GEMM count & Reconstruction \\
    \midrule
    AMX-FP32 & oneMKL SGEMM & 3 & 6 & FP32 \\
    AMX-FP64-6 & oneMKL DGEMM & 6 & 6 & FP64 \\
    AMX-FP64-10 & oneMKL DGEMM & 6 & 10 & FP64 \\
    AMX-FP64-15 & oneMKL DGEMM & 6 & 15 & FP64 \\
    AMX-FP64-21 & oneMKL DGEMM & 6 & 21 & FP64 \\
    \bottomrule
  \end{tabular}
  }
\end{table}

\subsection{Input Matrices and Accuracy Metrics}
\label{sec:inputs-metrics}

The numerical evaluation includes three representative classes of square input
matrices:

\begin{itemize}
  \item random matrices with uniformly distributed entries, used as the
  default non-adversarial case;
  \item signed log-uniform matrices, used to exercise a wider dynamic range
  while remaining inside $\FBsixtyfour$ for FP64 experiments; and
  \item cancellation-sensitive matrices, used to expose cases where product
  truncation and summation order have a larger effect.
\end{itemize}
The accuracy results reported in Section~\ref{sec:accuracy-results} use uniform
random inputs sampled from $[-1,1]$. Additional evaluations using positive
uniform, signed log-uniform, and cancellation-sensitive inputs are reported in
Appendix~\ref{app:supplementary-accuracy}.

Input data are generated directly in the precision of the path being tested.
The AMX-FP32 kernel and the SGEMM baseline use identical matrices generated
and stored in FP32, whereas all AMX-FP64 variants and their DGEMM reference use matrices
generated and stored directly in FP64 within the supported BF16 exponent
range. Thus, the FP64 operands retain FP64 significand information and are not
obtained by widening FP32 values.

All reported experiments use square GEMM problems with $m=n=k=N$, and the
matrix order $N$ ranges from 1024 to 32768 to expose small-matrix overheads
and large-matrix steady-state behavior. For FP32 accuracy evaluation, the original FP32 inputs
are exactly widened to FP64 and multiplied with oneMKL DGEMM. This FP64 result
is the common reference for AMX-FP32 and SGEMM, following the higher-precision
reference approach of Ootomo and Yokota~\cite{ootomo2022recovering}.
The FP32 outputs are widened before subtraction; the reference is not rounded
back to FP32. For AMX-FP64, the comparison reference remains oneMKL DGEMM
applied to the original FP64 inputs. These FP64-path comparisons measure
agreement with DGEMM, not error against an exact product.

For a computed result $\widehat C$ and the corresponding reference
$C_{\mathrm{ref}}$, we report the Frobenius-norm relative error
\begin{equation}
  e_F = \frac{\norm{\widehat C-C_{\mathrm{ref}}}_F}
  {\norm{C_{\mathrm{ref}}}_F},
  \label{eq:froerror}
\end{equation}
and the GEMM-scaled componentwise error
\begin{equation}
  e_{\mathrm{scaled}} =
  \max_{i,j}
  \frac{|(\widehat C-C_{\mathrm{ref}})_{ij}|}
  {(|A||B|)_{ij}},
  \label{eq:scalederror}
\end{equation}
where $|A||B|$ is evaluated in FP64 or a wider diagnostic format using the
original inputs. For FP32, both methods use the same denominator.
A zero denominator contributes zero when the corresponding discrepancy is
zero and infinity otherwise; absolute discrepancies are also checked.
The Frobenius-norm metric measures aggregate deviation from the reference,
whereas the scaled componentwise metric exposes the worst local discrepancy
relative to the absolute dot-product bound at the corresponding entry. The
claim of FP32-level accuracy is based on these two metrics rather than a single
scalar threshold, and it does not imply bitwise equality with oneMKL SGEMM.
This normwise and componentwise reporting follows established accuracy analyses
for mixed-precision and multiword matrix
multiplication~\cite{higham2002accuracy,fasi2023matrix}.

\subsection{Performance Metrics}

Performance is reported as wall-clock time $T$, effective throughput, and
speedup over the corresponding oneMKL baseline.  Because all reported
experiments use square GEMM with $m=n=k=N$, effective throughput is reported in
TFLOPS and computed as $2N^3/(T\times10^{12})$, where $T$ is the mean runtime per
execution in seconds.
This normalization counts the mathematical work of the requested
high-precision GEMM, not the larger number of BF16 component operations issued
by the emulation kernels.  Accuracy and performance are reported together for
every evaluated variant to make the accuracy-performance tradeoff explicit.
Timing excludes matrix initialization and reference-result generation;
one-time setup costs such as AMX permission checks and tile-configuration setup
are kept outside the steady-state timing region unless explicitly reported.
The performance evaluation reports full-core results and AMX-FP32 thread
scaling.

\section{Results}
\label{sec:results}

We first quantify the raw AMX-BF16 headroom, then report accuracy, performance,
and runtime composition for the implemented variants.

\subsection{AMX-BF16 Performance Headroom}
\label{sec:amx-headroom}

Before evaluating the emulation kernels, we measure the performance of the
oneMKL AMX-BF16 matrix-multiply routine, denoted here by the shorthand
\texttt{bf16bf16fp32}, which computes BF16-by-BF16 products with FP32
accumulation. Figure~\ref{fig:headroom} compares this routine with oneMKL SGEMM
and DGEMM on square matrices with $N=1024$ to $32768$, using 64 threads and the
affinity and timing protocol in Section~\ref{sec:methodology}. Two BF16 timings
are reported: compute only, with input packing excluded from the timed region,
and packing plus compute, with input packing included in each timed execution.
Neither includes the decomposition and reconstruction required for higher-precision
emulation. These measurements therefore provide library-level performance
references, not accuracy-equivalent alternatives to SGEMM or DGEMM.

\begin{figure}[H]
  \centering
  \hspace*{0.45cm}\begin{tikzpicture}[trim axis left, trim axis right]
    \begin{axis}[
      width=0.95\textwidth,
      height=7.0cm,
      ybar,
      bar width=10.0pt,
      enlarge x limits=0.14,
      ymin=0,
      ymax=58,
      ylabel={TFLOPS},
      ylabel style={font=\small, at={(axis description cs:-0.043,0.5)}, anchor=south},
      xlabel={Matrix order $N$},
      symbolic x coords={1024,2048,4096,8192,16384,32768},
      xtick=data,
      x tick label style={font=\scriptsize, rotate=35, anchor=east},
      ymajorgrids=true,
      grid style={draw=gray!20},
      axis line style={draw=amxnavy!70},
      tick style={draw=amxnavy!70},
      legend style={
        at={(0.5,1.08)},
        anchor=south,
        legend columns=2,
        cells={anchor=west},
        draw=none,
        font=\footnotesize
      },
      legend image code/.code={
        \draw[#1,draw=none] (0cm,-0.08cm) rectangle (0.18cm,0.08cm);
      },
      nodes near coords={\textbf{\pgfmathprintnumber[fixed,precision=2]{\pgfplotspointmeta}}},
      every node near coord/.append style={
        font=\scriptsize,
        text=amxnavy!85,
        rotate=90,
        anchor=west
      },
      every axis plot/.append style={draw=none}
    ]
      \addplot+[fill=headroomblue!78] coordinates {
        (1024,24.54267026) (2048,43.93828436) (4096,47.8881371)
        (8192,43.11810305) (16384,40.91206057) (32768,34.6644066)
      };
      \addlegendentry{BF16 (compute only)}

      \addplot+[fill=headroomyellow!85] coordinates {
        (1024,9.608427955) (2048,24.38936568) (4096,29.1802449)
        (8192,37.78390473) (16384,39.44436333) (32768,34.32621667)
      };
      \addlegendentry{BF16 (packing + compute)}

      \addplot+[fill=headroomgreen!72] coordinates {
        (1024,6.567228281) (2048,10.046707125) (4096,9.41362695)
        (8192,9.478548515) (16384,9.09626993) (32768,8.613065383)
      };
      \addlegendentry{SGEMM}

      \addplot+[fill=headroombrown!72] coordinates {
        (1024,4.194304) (2048,3.843371182) (4096,4.627574191)
        (8192,4.759790596) (16384,4.343749641) (32768,4.672559374)
      };
      \addlegendentry{DGEMM}
    \end{axis}
  \end{tikzpicture}
  \caption{oneMKL GEMM throughput on square matrices with 64 threads.
  BF16 denotes \texttt{bf16bf16fp32}, measured with and without input packing.
  Effective throughput is computed from $2N^3$ operations and the mean runtime;
  values above the bars are in TFLOPS.}
  \label{fig:headroom}
\end{figure}

Even with packing excluded, BF16 throughput is only $3.74$--$5.09\times$
that of SGEMM and $5.85$--$11.43\times$ that of DGEMM across the tested sizes.
These advantages leave limited headroom for a straightforward emulation based
on independent library calls: AMX-FP32 requires six component products,
while AMX-FP64 requires 6, 10, 15, or 21, in addition to decomposition and
reconstruction. Including packing further reduces the throughput ratios to
$1.46$--$4.34\times$ and $2.29$--$9.08\times$, respectively. The reduction is
particularly pronounced at the smaller tested sizes; at $N=1024$, the ratios
fall from $3.74\times$ to $1.46\times$ relative to SGEMM and from $5.85\times$
to $2.29\times$ relative to DGEMM. The relative impact of packing diminishes
as matrix size increases, consistent with its $O(N^2)$ work compared with the
$O(N^3)$ matrix multiplication.

Achieving end-to-end speedups over oneMKL therefore requires optimizations
beyond a sequence of independent BF16 GEMM calls. Reuse of packed inputs,
scheduling that reduces repeated data movement, and efficient reconstruction
are needed to improve the attainable performance of AMX-FP32 and AMX-FP64.
The AMX-FP64 path must additionally account for storing, widening, and
accumulating component results in FP64, which motivates evaluating the
retained product count jointly with accuracy. The library measurements are
a performance reference rather than an upper bound on the custom
implementations, whose data movement and accumulation costs can differ from
those of separate BF16 GEMM calls.

\subsection{Numerical Accuracy}
\label{sec:accuracy-results}

Tables~\ref{tab:accuracy} and~\ref{tab:accuracy-fp64} report the metrics defined
in Section~\ref{sec:inputs-metrics} for uniform random inputs sampled from
$[-1,1]$, with $N=1024$--$32768$. AMX-FP32 and SGEMM share an FP64 reference,
whereas the AMX-FP64 results quantify discrepancies from DGEMM on FP64 inputs.

\begin{table}[H]
  \centering
  \caption{FP32 numerical accuracy for square matrices with entries sampled from $U(-1,1)$.
  AMX-FP32 and oneMKL SGEMM are evaluated against the same FP64 DGEMM reference
  computed from the original FP32 inputs exactly widened to FP64.}
  \label{tab:accuracy}
  \small
  \setlength{\tabcolsep}{5.0pt}
  \renewcommand{\arraystretch}{1.15}
  \begin{tabular}{@{}c cc cc@{}}
    \toprule
    \multirow{2}{*}{$N$}
    & \multicolumn{2}{c}{AMX-FP32} & \multicolumn{2}{c}{oneMKL SGEMM} \\
    \cmidrule(lr){2-3}\cmidrule(lr){4-5}
    & $e_F$ & $e_{\mathrm{scaled}}$ & $e_F$ & $e_{\mathrm{scaled}}$ \\
    \midrule
    1024 & $1.08\times10^{-7}$ & $3.68\times10^{-8}$ & $3.38\times10^{-7}$ & $1.21\times10^{-7}$ \\
    2048 & $1.39\times10^{-7}$ & $2.91\times10^{-8}$ & $3.48\times10^{-7}$ & $9.45\times10^{-8}$ \\
    4096 & $1.49\times10^{-7}$ & $2.51\times10^{-8}$ & $3.53\times10^{-7}$ & $5.44\times10^{-8}$ \\
    8192 & $1.66\times10^{-7}$ & $2.27\times10^{-8}$ & $3.60\times10^{-7}$ & $3.89\times10^{-8}$ \\
    16384 & $1.95\times10^{-7}$ & $2.56\times10^{-8}$ & $3.71\times10^{-7}$ & $3.09\times10^{-8}$ \\
    32768 & $2.42\times10^{-7}$ & $2.56\times10^{-8}$ & $3.89\times10^{-7}$ & $4.42\times10^{-8}$ \\
    \bottomrule
  \end{tabular}
\end{table}

Against the common FP64 reference, AMX-FP32 gives
$e_F=(1.08$--$2.42)\times10^{-7}$, compared with
$(3.38$--$3.89)\times10^{-7}$ for SGEMM, and has lower scaled errors at every
tested order. These results support the six-product reconstruction as a means
of obtaining accuracy comparable to SGEMM from BF16 component products.
Residual-component products recover information discarded by a single BF16
conversion, while the final error also depends on the retained terms and their
accumulation \cite{fasi2023matrix}. The measured errors reflect their combined
effect and do not isolate the contribution of the accumulation schedule.

AMX-FP32's $e_F$ increases with $N$, whereas its scaled error remains within
$2.27\times10^{-8}$--$3.68\times10^{-8}$. These metrics need not follow the
same trend: signed products can cancel in the reference result used to
normalize $e_F$, but not in the absolute dot-product sums used for
$e_{\mathrm{scaled}}$. The additional distributions in
\ref{app:supplementary-accuracy} examine the sensitivity to input structure.

\begin{table}[H]
  \centering
  \caption{AMX-FP64 discrepancies relative to oneMKL DGEMM for square matrices with
  FP64 entries sampled from $U(-1,1)$. Both implementations use the same FP64 inputs.
  Entries use scientific e notation: \textnormal{1.49e-08} denotes
  $1.49\times10^{-8}$.}
  \label{tab:accuracy-fp64}
  \footnotesize
  \setlength{\tabcolsep}{3.5pt}
  \renewcommand{\arraystretch}{1.15}
  \begin{tabular}{@{}r rr rr rr rr@{}}
    \toprule
    \multirow{2}{*}{$N$}
    & \multicolumn{2}{c}{AMX-FP64-6} & \multicolumn{2}{c}{AMX-FP64-10}
    & \multicolumn{2}{c}{AMX-FP64-15} & \multicolumn{2}{c}{AMX-FP64-21} \\
    \cmidrule(lr){2-3}\cmidrule(lr){4-5}\cmidrule(lr){6-7}\cmidrule(lr){8-9}
    & $e_F$ & $e_{\mathrm{scaled}}$ & $e_F$ & $e_{\mathrm{scaled}}$
    & $e_F$ & $e_{\mathrm{scaled}}$ & $e_F$ & $e_{\mathrm{scaled}}$ \\
    \midrule
    1024 & \textnormal{1.49e-08} & \textnormal{3.38e-09} & \textnormal{3.26e-11} & \textnormal{6.93e-12} & \textnormal{6.97e-14} & \textnormal{1.40e-14} & \textnormal{6.96e-16} & \textnormal{2.79e-16} \\
    2048 & \textnormal{1.49e-08} & \textnormal{2.34e-09} & \textnormal{3.25e-11} & \textnormal{5.10e-12} & \textnormal{6.95e-14} & \textnormal{1.08e-14} & \textnormal{7.45e-16} & \textnormal{2.16e-16} \\
    4096 & \textnormal{1.49e-08} & \textnormal{1.73e-09} & \textnormal{3.25e-11} & \textnormal{3.87e-12} & \textnormal{6.96e-14} & \textnormal{7.65e-15} & \textnormal{8.19e-16} & \textnormal{1.67e-16} \\
    8192 & \textnormal{1.49e-08} & \textnormal{1.24e-09} & \textnormal{3.25e-11} & \textnormal{2.86e-12} & \textnormal{6.96e-14} & \textnormal{6.15e-15} & \textnormal{9.41e-16} & \textnormal{1.52e-16} \\
    16384 & \textnormal{1.49e-08} & \textnormal{9.36e-10} & \textnormal{3.25e-11} & \textnormal{2.10e-12} & \textnormal{6.96e-14} & \textnormal{4.40e-15} & \textnormal{1.14e-15} & \textnormal{1.61e-16} \\
    32768 & \textnormal{1.49e-08} & \textnormal{6.80e-10} & \textnormal{3.25e-11} & \textnormal{1.48e-12} & \textnormal{6.96e-14} & \textnormal{3.04e-15} & \textnormal{1.46e-15} & \textnormal{1.93e-16} \\
    \bottomrule
  \end{tabular}
\end{table}

For AMX-FP64, retaining more component products reduces both discrepancies.
The 6-, 10-, and 15-product variants give $e_F$ of approximately
$1.49\times10^{-8}$, $3.25\times10^{-11}$, and $6.96\times10^{-14}$,
respectively. The 21-product variant reaches
$6.96\times10^{-16}$--$1.46\times10^{-15}$ relative to DGEMM.
These measurements characterize the accuracy-cost tradeoff, with closer agreement with DGEMM as
more component products are retained.

\subsection{GEMM Performance}

This section reports GEMM performance under the multicore configuration
described in Section~\ref{sec:methodology}.  Unlike the headroom
study in Section~\ref{sec:amx-headroom}, these measurements include BF16
decomposition, packed component-buffer generation, AMX component products,
and target-format reconstruction.

\begin{figure}[!htb]
  \centering
  \hspace*{0.18cm}\begin{tikzpicture}[trim axis left, trim axis right]
    \begin{axis}[
      width=0.97\textwidth,
      height=6.0cm,
      ymin=0,
      ymax=13,
      enlarge x limits=0.07,
      ylabel={TFLOPS},
      ylabel style={font=\small, at={(axis description cs:-0.065,0.5)}, anchor=south},
      xlabel={Matrix order $N$},
      xlabel style={font=\small},
      symbolic x coords={1024,2048,4096,8192,16384,32768},
      xtick=data,
      x tick label style={font=\small, yshift=-0.25em},
      ymajorgrids=true,
      grid style={draw=gray!20},
      axis line style={draw=amxnavy!70},
      tick style={draw=amxnavy!70},
      legend style={
        at={(0.5,1.04)},
        anchor=south,
        draw=none,
        font=\footnotesize
      },
      legend cell align={left},
      every axis plot/.append style={line width=1.05pt, mark size=2.0pt}
    ]
      \addplot+[color=headroomblue, solid, mark=*,
        mark options={fill=headroomblue, draw=headroomblue}] coordinates {
        (1024,4.436949686) (2048,7.310582631) (4096,9.67879954)
        (8192,10.99511628) (16384,11.71250735) (32768,11.56809866)
      };
      \addlegendentry{AMX-FP32 (with scheduling optimization)}
      \addplot+[color=headroombrown, solid, mark=triangle*, mark size=3.5pt,
        mark options={solid, line width=0.5pt, fill=headroombrown, draw=headroombrown}] coordinates {
        (1024,3.6835054) (2048,4.655791107) (4096,5.898667531)
        (8192,5.581277298) (16384,5.353678042) (32768,5.275018304)
      };
      \addlegendentry{AMX-FP32 (without scheduling optimization)}
      \addplot+[color=headroomgreen!85!black, dashed, mark=square*,
        mark options={solid, fill=headroomgreen!85!black, draw=headroomgreen!85!black}] coordinates {
        (1024,6.567228281) (2048,10.046707125) (4096,9.41362695)
        (8192,9.478548515) (16384,9.09626993) (32768,8.613065383)
      };
      \addlegendentry{MKL SGEMM}
    \end{axis}
  \end{tikzpicture}

  \vspace{0.6em}

  \hspace*{0.18cm}\begin{tikzpicture}[trim axis left, trim axis right]
    \begin{axis}[
      width=0.97\textwidth,
      height=4.5cm,
      ymin=0,
      ymax=1.7,
      enlarge x limits=0.07,
      ylabel={Speedup},
      ylabel style={font=\small, at={(axis description cs:-0.065,0.5)}, anchor=south},
      xlabel={Matrix order $N$},
      xlabel style={font=\small},
      symbolic x coords={1024,2048,4096,8192,16384,32768},
      xtick=data,
      x tick label style={font=\small, yshift=-0.25em},
      ymajorgrids=true,
      grid style={draw=gray!20},
      axis line style={draw=amxnavy!70},
      tick style={draw=amxnavy!70},
      every axis plot/.append style={line width=1.05pt, mark size=2.0pt}
    ]
      \draw[gray!70, densely dotted] (rel axis cs:0,0.58823529) --
        (rel axis cs:1,0.58823529);
      \addplot+[color=headroomblue, mark=*,
        mark options={fill=headroomblue, draw=headroomblue},
        nodes near coords={\pgfmathprintnumber[fixed,precision=2]{\pgfplotspointmeta}},
        every node near coord/.append style={font=\scriptsize, anchor=south, yshift=2pt}] coordinates {
        (1024,0.67562) (2048,0.727659574) (4096,1.028169014)
        (8192,1.16) (16384,1.28762) (32768,1.34309)
      };
      \addplot+[color=headroombrown, dashed, mark=triangle*, mark size=3.5pt,
        mark options={solid, line width=0.5pt, fill=headroombrown, draw=headroombrown},
        nodes near coords={\pgfmathprintnumber[fixed,precision=2]{\pgfplotspointmeta}},
        every node near coord/.append style={font=\scriptsize, anchor=north, yshift=-4pt}] coordinates {
        (1024,0.56089) (2048,0.463414634) (4096,0.626609442)
        (8192,0.58883) (16384,0.58856) (32768,0.61244)
      };
    \end{axis}
  \end{tikzpicture}
  \caption{AMX-FP32 performance with and without scheduling optimization
  against oneMKL SGEMM on 64 threads. The upper panel reports effective
  throughput in TFLOPS; the lower panel reports the total-runtime ratio
  $T_{\mathrm{SGEMM}}/T_{\mathrm{AMX\text{-}FP32}}$ for each variant, using the
  same colors and markers. The dotted line marks equal runtime; ratios above
  one indicate a speedup over SGEMM. Both AMX-FP32 timings include decomposition,
  packing, component products, and reconstruction.}
  \label{fig:fullcore-fp32}
\end{figure}

Figure~\ref{fig:fullcore-fp32} compares AMX-FP32 with and without the
operand-reuse scheduling optimization against oneMKL SGEMM. Both AMX-FP32
variants use the six-product construction.
The optimized variant is slower than SGEMM at $N=1024$ and $2048$, with speedup
ratios of $0.68\times$ and $0.73\times$, respectively. At these smaller orders,
decomposition and packing account for a substantial fraction of total runtime,
limiting the overall benefit of AMX acceleration, as shown by the runtime
breakdown in Figure~\ref{fig:breakdown}. As matrix order increases, the
$O(N^2)$ preprocessing work is amortized over the $O(N^3)$ component
computation, reducing its relative overhead. The optimized variant's measured
speedup increases to $1.03\times$ at $N=4096$ and $1.16$--$1.34\times$ for
$N\ge8192$. Without the scheduling optimization, AMX-FP32 remains slower than
SGEMM throughout the tested range, with ratios of $0.46$--$0.63\times$.
The optimized variant is $1.20$--$2.19\times$ faster than the variant without
scheduling optimization, demonstrating the importance of scheduling to the
end-to-end performance of this implementation.

These results are consistent with the limited BF16 performance headroom in
Figure~\ref{fig:headroom}. The operand-reuse schedule reduces repeated tile
loads and retains partial sums in FP32
accumulator tiles across component products, enabling gains beyond a
straightforward execution of separate products.

\begin{figure}[!htb]
  \centering
  \hspace*{-0.85cm}\begin{tikzpicture}
    \begin{groupplot}[
      group style={group size=3 by 1, horizontal sep=1.03cm},
      width=0.345\textwidth,
      height=5.25cm,
      ymin=0,
      enlarge x limits=0.07,
      symbolic x coords={2,4,8,16,32,64},
      xtick=data,
      x tick label style={font=\scriptsize},
      xlabel={Threads},
      xlabel style={font=\scriptsize},
      ymajorgrids=true,
      grid style={draw=gray!20},
      axis line style={draw=amxnavy!70},
      tick style={draw=amxnavy!70},
      legend style={
        at={(1.72,1.17)},
        anchor=south,
        legend columns=2,
        draw=none,
        font=\scriptsize
      },
      legend cell align={left},
      every axis plot/.append style={line width=1.05pt, mark size=2.0pt}
    ]
      \nextgroupplot[
        title={$N=1024$},
        title style={font=\small},
        ylabel={TFLOPS},
        ylabel style={font=\small, at={(axis description cs:-0.155,0.5)}, anchor=south},
        ymax=7.5
      ]
      \addplot+[color=headroomblue, mark=*,
        mark options={fill=headroomblue, draw=headroomblue}] coordinates {
        (2,0.482580595) (4,0.786624047) (8,1.227133513) (16,1.481023206) (32,3.153426796) (64,4.436949686)
      };
      \addlegendentry{AMX-FP32}
      \addplot+[color=black!62, mark=square*,
        mark options={fill=black!62, draw=black!62}] coordinates {
        (2,0.370895276) (4,0.644889984) (8,1.112685828) (16,2.565691336) (32,4.559413265) (64,6.567228281)
      };
      \addlegendentry{MKL SGEMM}

      \nextgroupplot[
        title={$N=2048$},
        title style={font=\small},
        ymax=11.5
      ]
      \addplot+[color=headroomblue, mark=*,
        mark options={fill=headroomblue, draw=headroomblue}] coordinates {
        (2,0.479884614) (4,0.928641578) (8,1.769296517) (16,2.696996732) (32,5.082801534) (64,7.310582631)
      };
      \addplot+[color=black!62, mark=square*,
        mark options={fill=black!62, draw=black!62}] coordinates {
        (2,0.377579543) (4,0.679046213) (8,1.352745605) (16,2.180186445) (32,5.436667463) (64,10.046707125)
      };

      \nextgroupplot[
        title={$N=4096$},
        title style={font=\small},
        ymax=11.5
      ]
      \addplot+[color=headroomblue, mark=*,
        mark options={fill=headroomblue, draw=headroomblue}] coordinates {
        (2,0.509033161) (4,1.010580540) (8,1.880149842) (16,3.145056144) (32,6.135667566) (64,9.678799540)
      };
      \addplot+[color=black!62, mark=square*,
        mark options={fill=black!62, draw=black!62}] coordinates {
        (2,0.368469044) (4,0.683775888) (8,1.405306273) (16,2.494354872) (32,5.327091220) (64,9.413626950)
      };

    \end{groupplot}
  \end{tikzpicture}

  \vspace{0.6em}

  \hspace*{0.18cm}\begin{tikzpicture}[trim axis left, trim axis right]
    \begin{axis}[
      width=0.97\textwidth,
      height=4.5cm,
      ymin=0.4,
      ymax=1.65,
      enlarge x limits=0.07,
      ylabel={Speedup},
      ylabel style={font=\small, at={(axis description cs:-0.075,0.5)}, anchor=south},
      xlabel={OpenMP threads},
      xlabel style={font=\small},
      symbolic x coords={2,4,8,16,32,64},
      xtick=data,
      x tick label style={font=\small, yshift=-0.25em},
      ymajorgrids=true,
      ytick={0.5,0.75,1,1.25,1.5},
      grid style={draw=gray!20},
      axis line style={draw=amxnavy!70},
      tick style={draw=amxnavy!70},
      legend style={
        at={(0.5,1.05)},
        anchor=south,
        legend columns=3,
        draw=none,
        font=\scriptsize
      },
      legend cell align={left},
      every axis plot/.append style={line width=1.05pt, mark size=2.1pt}
    ]
      \addplot[black!65, densely dashed, no marks, forget plot,
        line width=1.1pt] coordinates {(2,1) (64,1)};
      \addplot+[color=headroomblue, mark=*,
        mark options={solid, fill=headroomblue, draw=headroomblue}] coordinates {
        (2,1.301123596) (4,1.219780220) (8,1.102857143) (16,0.577241379) (32,0.691629956) (64,0.675619835)
      };
      \addlegendentry{$N=1024$}
      \addplot+[color=headroomgreen!85!black, mark=square*,
        mark options={solid, fill=headroomgreen!85!black, draw=headroomgreen!85!black}] coordinates {
        (2,1.270949721) (4,1.367567568) (8,1.307929969) (16,1.237048666) (32,0.934911243) (64,0.727659574)
      };
      \addlegendentry{$N=2048$}
      \addplot+[color=headroombrown, mark=triangle*, mark size=3.2pt,
        mark options={solid, fill=headroombrown, draw=headroombrown}] coordinates {
        (2,1.381481481) (4,1.477941176) (8,1.337893297) (16,1.260869565) (32,1.151785714) (64,1.028169014)
      };
      \addlegendentry{$N=4096$}
    \end{axis}
  \end{tikzpicture}
  \caption{Thread scaling of optimized AMX-FP32 against oneMKL SGEMM for
  $N=1024$, $2048$, and $4096$. The upper panels report effective throughput
  in TFLOPS; the lower panel reports the total-runtime ratio
  $T_{\mathrm{SGEMM}}/T_{\mathrm{AMX\text{-}FP32}}$. The dashed line marks equal
  runtime. AMX-FP32 timings include decomposition, packing, component products,
  and reconstruction.}
  \label{fig:fp32-scaling}
\end{figure}

Figure~\ref{fig:fp32-scaling} focuses on AMX-FP32 for smaller square
matrices ($N=1024$, $2048$, and $4096$) with limited threads per GEMM.
This configuration is relevant to HPC applications that distribute work across
processes or independent tasks, with each local matrix multiplication using
only a subset of the available cores. At $N=1024$, AMX-FP32 outperforms
oneMKL SGEMM by $1.10$--$1.30\times$ with 2--8 threads. At $N=2048$,
the advantage extends to 2--16 threads, with speedups of
$1.24$--$1.37\times$. For $N=4096$, speedups of $1.26$--$1.48\times$
are obtained with 2--16 threads. For $N=1024$, SGEMM scales more strongly
between 8 and 16 threads, reversing the performance ordering. These results
demonstrate that AMX-FP32 can also benefit smaller matrix multiplications
when each operation is assigned a modest thread count.

Profiling shows that the fraction of total runtime spent on decomposition
and packing decreases substantially as the thread count is reduced,
reflecting unequal scaling of preprocessing and component computation.
Decomposition and packing involve elementwise conversions and data rearrangement
with limited reuse; their scaling can be constrained by memory bandwidth,
work granularity, and parallel overhead. The blocked component GEMMs perform
more arithmetic per data access and can benefit more from additional cores.
Reducing the thread count can therefore increase component-computation time
proportionally more than preprocessing time, lowering the preprocessing
fraction even if its absolute time increases. The relative speedups over SGEMM
at lower thread counts reflect the combined effects of AMX's arithmetic
throughput, preprocessing overhead, and the thread-scaling behavior of both
implementations. Figure~\ref{fig:breakdown} examines the decomposition and
packing fractions at 64 threads.

\begin{figure}[!htb]
  \centering
  \hspace*{0.18cm}\begin{tikzpicture}[trim axis left, trim axis right]
    \begin{axis}[
      width=0.97\textwidth,
      height=6.2cm,
      ymin=0,
      ymax=9,
      enlarge x limits=0.07,
      ylabel={TFLOPS},
      ylabel style={font=\small, at={(axis description cs:-0.065,0.5)}, anchor=south},
      xlabel={Matrix order $N$},
      xlabel style={font=\small},
      symbolic x coords={1024,2048,4096,8192,16384,32768},
      xtick=data,
      x tick label style={font=\small, yshift=-0.25em},
      ymajorgrids=true,
      grid style={draw=gray!20},
      axis line style={draw=amxnavy!70},
      tick style={draw=amxnavy!70},
      legend style={
        at={(0.02,0.98)},
        anchor=north west,
        draw=none,
        fill=white,
        fill opacity=0.85,
        text opacity=1,
        font=\scriptsize
      },
      legend cell align={left},
      every axis plot/.append style={line width=1.05pt, mark size=2.35pt}
    ]
      \addplot+[color=headroomblue, mark=*,
        mark options={solid, fill=headroomblue, draw=headroomblue}] coordinates {
        (1024,0.998829604) (2048,1.881694325) (4096,4.755673131) (8192,6.915167470) (16384,7.518028224) (32768,7.724340744)
      };
      \addlegendentry{AMX-FP64-6}
      \addplot+[color=amxgreen!75!black, mark=square*,
        mark options={solid, fill=amxgreen!75!black, draw=amxgreen!75!black}] coordinates {
        (1024,0.883738127) (2048,1.605595251) (4096,3.569842947) (8192,4.543436478) (16384,5.143914048) (32768,5.095491975)
      };
      \addlegendentry{AMX-FP64-10}
      \addplot+[color=amxorange!90!black, mark=triangle*, mark size=3.2pt,
        mark options={solid, fill=amxorange!90!black, draw=amxorange!90!black}] coordinates {
        (1024,0.789516047) (2048,1.396737332) (4096,2.689607700) (8192,3.205573259) (16384,3.306801888) (32768,3.311470314)
      };
      \addlegendentry{AMX-FP64-15}
      \addplot+[color=headroombrown, mark=diamond*,
        mark options={solid, fill=headroombrown, draw=headroombrown}] coordinates {
        (1024,0.699506074) (2048,1.184818564) (4096,2.063647950) (8192,2.334419592) (16384,2.360100086) (32768,2.357411865)
      };
      \addlegendentry{AMX-FP64-21}
      \addplot+[color=black!78, dashed, line width=1.35pt, mark=pentagon*,
        mark size=3.1pt,
        mark options={solid, fill=black, draw=black, line width=0.9pt}] coordinates {
        (1024,4.194304000) (2048,3.843371182) (4096,4.627574191) (8192,4.759790596) (16384,4.343749641) (32768,4.672559374)
      };
      \addlegendentry{MKL DGEMM}
    \end{axis}
  \end{tikzpicture}

  \vspace{0.6em}

  \hspace*{0.18cm}\begin{tikzpicture}[trim axis left, trim axis right]
    \begin{axis}[
      width=0.97\textwidth,
      height=4.5cm,
      ymin=0,
      ymax=1.9,
      enlarge x limits=0.07,
      ylabel={Speedup},
      ylabel style={font=\small, at={(axis description cs:-0.065,0.5)}, anchor=south},
      xlabel={Matrix order $N$},
      xlabel style={font=\small},
      symbolic x coords={1024,2048,4096,8192,16384,32768},
      xtick=data,
      x tick label style={font=\small, yshift=-0.25em},
      ymajorgrids=true,
      grid style={draw=gray!20},
      axis line style={draw=amxnavy!70},
      tick style={draw=amxnavy!70},
      legend style={
        at={(0.5,1.06)},
        anchor=south,
        legend columns=4,
        draw=none,
        font=\scriptsize
      },
      legend cell align={left},
      every axis plot/.append style={line width=1.05pt, mark size=2.1pt}
    ]
      \addplot[color=black!65, densely dashed, no marks, line width=1.1pt, forget plot]
        coordinates {(1024,1) (32768,1)};
      \addplot+[color=headroomblue, mark=*,
        mark options={solid, fill=headroomblue, draw=headroomblue}] coordinates {
        (1024,0.238139535) (2048,0.489594743) (4096,1.027681661) (8192,1.452830189) (16384,1.730769231) (32768,1.653128430)
      };
      \addlegendentry{AMX-FP64-6}
      \addplot+[color=amxgreen!75!black, mark=square*,
        mark options={solid, fill=amxgreen!75!black, draw=amxgreen!75!black}] coordinates {
        (1024,0.210699588) (2048,0.417757009) (4096,0.771428571) (8192,0.954545455) (16384,1.184210526) (32768,1.090514120)
      };
      \addlegendentry{AMX-FP64-10}
      \addplot+[color=amxorange!90!black, mark=triangle*, mark size=3.2pt,
        mark options={solid, fill=amxorange!90!black, draw=amxorange!90!black}] coordinates {
        (1024,0.188235294) (2048,0.363414634) (4096,0.581213307) (8192,0.673469388) (16384,0.761278195) (32768,0.708705882)
      };
      \addlegendentry{AMX-FP64-15}
      \addplot+[color=headroombrown, mark=diamond*,
        mark options={solid, fill=headroombrown, draw=headroombrown}] coordinates {
        (1024,0.166775244) (2048,0.308275862) (4096,0.445945946) (8192,0.490445860) (16384,0.543332439) (32768,0.504522613)
      };
      \addlegendentry{AMX-FP64-21}
    \end{axis}
  \end{tikzpicture}
  \caption{AMX-FP64 performance against oneMKL DGEMM on 64 threads for square
  matrices with $N=1024$--$32768$. The upper panel reports effective
  throughput in TFLOPS; the lower panel reports the total-runtime ratio
  $T_{\mathrm{DGEMM}}/T_{\mathrm{AMX\text{-}FP64}}$ for each retained
  product count. The dashed horizontal line marks equal runtime.
  All AMX-FP64 timings include six-slice decomposition, packing, component
  products, and FP64 reconstruction.}
  \label{fig:fullcore-fp64}
\end{figure}

Figure~\ref{fig:fullcore-fp64} compares the four AMX-FP64 variants with
oneMKL DGEMM on 64 threads, including decomposition, packing, component
products, and FP64 reconstruction in each AMX timing.
All four variants are slower than DGEMM at $N=1024$ and $2048$.
AMX-FP64-6 first exceeds DGEMM at the tested order $N=4096$, with a
measured speedup of $1.03\times$, and achieves $1.45$--$1.73\times$
for $N=8192$--$32768$. AMX-FP64-10 remains below DGEMM through $N=8192$,
but reaches $1.18\times$ and $1.09\times$ at $N=16384$ and $32768$,
respectively. AMX-FP64-15 and AMX-FP64-21 do not outperform DGEMM at any
tested order.

The results illustrate the limited performance headroom for FP64 emulation
identified in Figure~\ref{fig:headroom}. All four variants generate and pack
six BF16 slices per input matrix, so retaining fewer products does not remove
this common preprocessing work. As with AMX-FP32, preprocessing accounts for
a smaller fraction of total runtime as matrix order increases.
However, increasing the retained product count adds both BF16 GEMMs and
FP64 reconstruction work. Unlike AMX-FP32, each AMX-FP64 component result
must be stored, widened, and accumulated outside the AMX tiles.
The ten-product variant consequently has a smaller performance margin,
while retaining 15 or 21 products consumes the available advantage in these
measurements. Thus, the retained product count must be selected jointly
with the required accuracy: the speedups of variants retaining fewer products should
be interpreted alongside the numerical errors in
Section~\ref{sec:accuracy-results}, not as evidence of DGEMM-equivalent accuracy.
As with AMX-FP32, reducing the thread count can further improve AMX-FP64
performance relative to DGEMM.

\FloatBarrier
\subsection{Runtime Breakdown}

Figure~\ref{fig:breakdown} shows the fractions of total runtime spent on
preprocessing (BF16 decomposition and packing) and the computation stage,
which includes reconstruction, for each AMX implementation on 64 threads.
For AMX-FP32, preprocessing
accounts for $61.0\%$ and $37.2\%$ of runtime at $N=1024$ and $2048$,
respectively, helping explain the lower performance relative to SGEMM in
Figure~\ref{fig:fullcore-fp32}. This is consistent with
Figure~\ref{fig:headroom}, where input preparation substantially narrows
the BF16 performance advantage at smaller orders.
At the two largest orders, the AMX-FP32 preprocessing fraction decreases
from $3.8\%$ at $N=16384$ to $1.9\%$ at $N=32768$.
The gains from scheduling optimization in Figure~\ref{fig:fullcore-fp32}
are therefore mainly associated with the computation stage.

The AMX-FP64 results expose a related limitation. At $N=32768$,
preprocessing accounts for only $1.2\%$--$4.0\%$ of total runtime across
the four variants, yet the 15- and 21-product variants remain slower
than DGEMM in Figure~\ref{fig:fullcore-fp64}.
All four variants share the same six-slice preprocessing. The additional
component products retained to improve numerical accuracy increase the work
in the computation stage, including FP64 reconstruction. This helps explain
why the 15- and 21-product variants remain slower than DGEMM despite their
small preprocessing fractions.

\begin{figure}[H]
  \centering
  \begin{minipage}[t]{0.065\textwidth}
    \centering\scriptsize\bfseries\strut $\boldsymbol{N}$\par\vspace{0.45em}
    \begin{tikzpicture}
      \path[use as bounding box] (0,0) rectangle (0.80cm,4.20cm);
      \foreach \n/\y in {1024/3.85,2048/3.15,4096/2.45,8192/1.75,16384/1.05,32768/0.35} {
        \node[anchor=east,font=\scriptsize] at (0.78cm,\y cm) {\n};
      }
    \end{tikzpicture}
  \end{minipage}\hfill
  \begin{minipage}[t]{0.177\textwidth}
    \centering\scriptsize\bfseries\strut AMX-FP32\par\vspace{0.45em}
    \begin{tikzpicture}[trim axis left, trim axis right]
      \begin{axis}[
        width=\linewidth, height=4.20cm, scale only axis, xbar stacked, xmin=0, xmax=100,
        xtick={0,50,100}, xticklabels={,,},
        after end axis/.code={
          \node[font=\scriptsize\normalfont, inner sep=0pt, anchor=north west] at (rel axis cs:0,-0.040) {0\%};
          \node[font=\scriptsize\normalfont, inner sep=0pt, anchor=north] at (rel axis cs:0.5,-0.040) {50\%};
          \node[font=\scriptsize\normalfont, inner sep=0pt, anchor=north east] at (rel axis cs:1,-0.040) {100\%};
        },
        y dir=reverse, symbolic y coords={1024,2048,4096,8192,16384,32768},
        ytick=\empty, axis lines=box, enlarge y limits=0.10, /tikz/bar width=7.4pt,
        xmajorgrids=true, grid style={draw=gray!18}, axis line style={draw=amxnavy!70},
        tick style={draw=amxnavy!70}, every axis plot/.append style={draw=none}
      ]
        \addplot+[fill=headroomblue] coordinates {(60.9504,1024) (37.1915,2048) (19.7183,4096) (13.4000,8192) (3.8216,16384) (1.8905,32768)};
        \addplot+[fill=amxgreen!62] coordinates {(39.0496,1024) (62.8085,2048) (80.2817,4096) (86.6000,8192) (96.1784,16384) (98.1095,32768)};
      \end{axis}
    \end{tikzpicture}
  \end{minipage}\hfill
  \begin{minipage}[t]{0.177\textwidth}
    \centering\scriptsize\bfseries\strut AMX-FP64-6\par\vspace{0.45em}
    \begin{tikzpicture}[trim axis left, trim axis right]
      \begin{axis}[
        width=\linewidth, height=4.20cm, scale only axis, xbar stacked, xmin=0, xmax=100,
        xtick={0,50,100}, xticklabels={,,},
        after end axis/.code={
          \node[font=\scriptsize\normalfont, inner sep=0pt, anchor=north west] at (rel axis cs:0,-0.040) {0\%};
          \node[font=\scriptsize\normalfont, inner sep=0pt, anchor=north] at (rel axis cs:0.5,-0.040) {50\%};
          \node[font=\scriptsize\normalfont, inner sep=0pt, anchor=north east] at (rel axis cs:1,-0.040) {100\%};
        },
        y dir=reverse, symbolic y coords={1024,2048,4096,8192,16384,32768},
        ytick=\empty, axis lines=box, enlarge y limits=0.10, /tikz/bar width=7.4pt,
        xmajorgrids=true, grid style={draw=gray!18}, axis line style={draw=amxnavy!70},
        tick style={draw=amxnavy!70}, every axis plot/.append style={draw=none}
      ]
        \addplot+[fill=headroomblue] coordinates {(82.3256,1024) (70.9748,2048) (42.2145,4096) (20.6918,8192) (9.0598,16384) (3.9846,32768)};
        \addplot+[fill=amxgreen!62] coordinates {(17.6744,1024) (29.0252,2048) (57.7855,4096) (79.3082,8192) (90.9402,16384) (96.0154,32768)};
      \end{axis}
    \end{tikzpicture}
  \end{minipage}\hfill
  \begin{minipage}[t]{0.177\textwidth}
    \centering\scriptsize\bfseries\strut AMX-FP64-10\par\vspace{0.45em}
    \begin{tikzpicture}[trim axis left, trim axis right]
      \begin{axis}[
        width=\linewidth, height=4.20cm, scale only axis, xbar stacked, xmin=0, xmax=100,
        xtick={0,50,100}, xticklabels={,,},
        after end axis/.code={
          \node[font=\scriptsize\normalfont, inner sep=0pt, anchor=north west] at (rel axis cs:0,-0.040) {0\%};
          \node[font=\scriptsize\normalfont, inner sep=0pt, anchor=north] at (rel axis cs:0.5,-0.040) {50\%};
          \node[font=\scriptsize\normalfont, inner sep=0pt, anchor=north east] at (rel axis cs:1,-0.040) {100\%};
        },
        y dir=reverse, symbolic y coords={1024,2048,4096,8192,16384,32768},
        ytick=\empty, axis lines=box, enlarge y limits=0.10, /tikz/bar width=7.4pt,
        xmajorgrids=true, grid style={draw=gray!18}, axis line style={draw=amxnavy!70},
        tick style={draw=amxnavy!70}, every axis plot/.append style={draw=none}
      ]
        \addplot+[fill=headroomblue] coordinates {(72.8395,1024) (60.5607,2048) (31.6883,4096) (13.5950,8192) (6.1988,16384) (2.6285,32768)};
        \addplot+[fill=amxgreen!62] coordinates {(27.1605,1024) (39.4393,2048) (68.3117,4096) (86.4050,8192) (93.8012,16384) (97.3715,32768)};
      \end{axis}
    \end{tikzpicture}
  \end{minipage}\hfill
  \begin{minipage}[t]{0.177\textwidth}
    \centering\scriptsize\bfseries\strut AMX-FP64-15\par\vspace{0.45em}
    \begin{tikzpicture}[trim axis left, trim axis right]
      \begin{axis}[
        width=\linewidth, height=4.20cm, scale only axis, xbar stacked, xmin=0, xmax=100,
        xtick={0,50,100}, xticklabels={,,},
        after end axis/.code={
          \node[font=\scriptsize\normalfont, inner sep=0pt, anchor=north west] at (rel axis cs:0,-0.040) {0\%};
          \node[font=\scriptsize\normalfont, inner sep=0pt, anchor=north] at (rel axis cs:0.5,-0.040) {50\%};
          \node[font=\scriptsize\normalfont, inner sep=0pt, anchor=north east] at (rel axis cs:1,-0.040) {100\%};
        },
        y dir=reverse, symbolic y coords={1024,2048,4096,8192,16384,32768},
        ytick=\empty, axis lines=box, enlarge y limits=0.10, /tikz/bar width=7.4pt,
        xmajorgrids=true, grid style={draw=gray!18}, axis line style={draw=amxnavy!70},
        tick style={draw=amxnavy!70}, every axis plot/.append style={draw=none}
      ]
        \addplot+[fill=headroomblue] coordinates {(65.0735,1024) (52.6829,2048) (23.8748,4096) (9.5918,8192) (3.9850,16384) (1.7082,32768)};
        \addplot+[fill=amxgreen!62] coordinates {(34.9265,1024) (47.3171,2048) (76.1252,4096) (90.4082,8192) (96.0150,16384) (98.2918,32768)};
      \end{axis}
    \end{tikzpicture}
  \end{minipage}\hfill
  \begin{minipage}[t]{0.177\textwidth}
    \centering\scriptsize\bfseries\strut AMX-FP64-21\par\vspace{0.45em}
    \begin{tikzpicture}[trim axis left, trim axis right]
      \begin{axis}[
        width=\linewidth, height=4.20cm, scale only axis, xbar stacked, xmin=0, xmax=100,
        xtick={0,50,100}, xticklabels={,,},
        after end axis/.code={
          \node[font=\scriptsize\normalfont, inner sep=0pt, anchor=north west] at (rel axis cs:0,-0.040) {0\%};
          \node[font=\scriptsize\normalfont, inner sep=0pt, anchor=north] at (rel axis cs:0.5,-0.040) {50\%};
          \node[font=\scriptsize\normalfont, inner sep=0pt, anchor=north east] at (rel axis cs:1,-0.040) {100\%};
        },
        y dir=reverse, symbolic y coords={1024,2048,4096,8192,16384,32768},
        ytick=\empty, axis lines=box, enlarge y limits=0.10, /tikz/bar width=7.4pt,
        xmajorgrids=true, grid style={draw=gray!18}, axis line style={draw=amxnavy!70},
        tick style={draw=amxnavy!70}, every axis plot/.append style={draw=none}
      ]
        \addplot+[fill=headroomblue] coordinates {(57.6547,1024) (44.6897,2048) (18.3183,4096) (6.9851,8192) (2.8441,16384) (1.2161,32768)};
        \addplot+[fill=amxgreen!62] coordinates {(42.3453,1024) (55.3103,2048) (81.6817,4096) (93.0149,8192) (97.1559,16384) (98.7839,32768)};
      \end{axis}
    \end{tikzpicture}
  \end{minipage}
  \par\vspace{0.35em}
  {\footnotesize
    \raisebox{0.15ex}{\tikz\fill[headroomblue] (0,0) rectangle (1.1em,1.1ex);}\ \!Preprocessing
    \qquad
    \raisebox{0.15ex}{\tikz\fill[amxgreen!62] (0,0) rectangle (1.1em,1.1ex);}\ \!Compute + reconstruction
  }
  \caption{Runtime breakdown on 64 threads for square matrices with
  $N=1024$--$32768$. Each horizontal bar sums to $100\%$ of the measured
  total runtime. Blue denotes preprocessing, comprising BF16 decomposition
  and packing (including VNNI packing of $B$); green denotes the computation
  stage, including target-format reconstruction. The shared $N$ column identifies
  matrix order.}
  \label{fig:breakdown}
\end{figure}

\section{Discussion and Future Work}
\label{sec:limitations}

This paper is an early algorithmic study of BF16-based high-precision GEMM on a
CPU matrix engine. Its evaluation is deliberately limited to real, square
$C=AB$ problems, so it does not establish performance for rectangular, skinny,
or more general matrix-multiplication interfaces. Future work will extend the
kernels and evaluation to these matrix shapes and more complete BLAS semantics,
with corresponding changes to blocking, packing reuse, and scheduling.
The AMX-FP64 variants offer an accuracy-performance tradeoff and do not
guarantee complete FP64 or DGEMM-equivalent accuracy.

The attainable speedup is also constrained by the current AMX tile-register
resources. The eight-tile register file must simultaneously hold BF16 source
panels and FP32 accumulator tiles, which limits the $32\times32$ microkernel,
the number of component panels that can remain resident, and the amount of tile
load and store traffic that can be eliminated by scheduling. If future CPU matrix
engines provide more tile registers, larger tile capacity, or both, the same
decomposition framework could use larger microkernels and retain more operands
or accumulators on chip. Such hardware evolution would enlarge the performance
headroom available to extended-precision algorithms; it would not require a
change to the central BF16 component-product formulation.

The component-expansion principle is also not specific to x86 AMX. The method
and implementation logic described here can be adapted to CPUs that integrate
high-throughput low-precision matrix hardware, including Arm SME and SME2
implementations such as the planned FUJITSU-MONAKA-X processor for FugakuNEXT
and the LingKun LX2 processor~\cite{riken2026fugakunext,wang2026lx2sme}. Such
adaptations require platform-specific packing, blocking, and microkernel design,
but retain the same component-decomposition and wider-accumulation principle.

Alongside these performance and portability opportunities, the unscaled
AMX-FP64 path has a narrower input domain than native DGEMM. The BF16 normal
exponent range matches FP32 and covers the operand ranges of most HPC cases
targeted by this work, particularly after the common normalization or
nondimensionalization of physical variables. It nevertheless does not cover the
full FP64 exponent range, and the current method intentionally omits stored
exponent scaling. In addition to this input-range restriction, all nonzero
extracted components must be exactly representable as normal BF16 values.
Inputs or components outside the supported domain, exceptional IEEE values,
and cases involving overflow or underflow are unsupported. Extending the
framework with scale-aware decomposition and fallback handling is therefore
a key future step toward a broader FP64 input domain and more general HPC
deployment.

\section{Conclusion}

This paper presents a framework that uses Intel AMX, specifically its AMX-BF16
tile instructions, to emulate real, square FP32 and FP64 matrix multiplication. The AMX-FP32
path combines a three-component BF16
residual decomposition with the six-product schedule suggested by Henry et al.
and an operand-reuse schedule that keeps FP32 AMX
accumulator tiles resident and reduces tile-load traffic. It targets FP32-level
accuracy comparable to oneMKL SGEMM. For
BF16-range FP64 inputs, the AMX-FP64 path uses a
simplified fixed six-slice Ozaki decomposition. Each selected BF16 component
product is materialized, converted from FP32, and accumulated in FP64; retaining
6, 10, 15, or 21 products controls the accuracy-performance tradeoff.

In the reported full-core square experiments, the optimized AMX-FP32
implementation outperforms oneMKL SGEMM by $1.03$--$1.34\times$ for tested
orders $N\ge4096$, but remains slower at $N=1024$ and $2048$.
Against a common FP64 reference, its errors are lower than SGEMM's on the
tested $U(-1,1)$ inputs, supporting FP32-level accuracy. Without scheduling optimization,
AMX-FP32 does not outperform SGEMM at any tested order.
For $N\in\{1024,2048,4096\}$, the optimized AMX-FP32 implementation
achieves speedups of $1.10$--$1.48\times$ over SGEMM with 2--8 threads.
The full-core AMX-FP64 results show a narrower performance
window: AMX-FP64-6 reaches up to $1.73\times$ over oneMKL DGEMM, whereas
AMX-FP64-10 provides only a small gain for the two largest tested matrices and
AMX-FP64-15 and AMX-FP64-21 do not outperform DGEMM. Retaining more
products increases both BF16 computation and FP64 reconstruction costs,
narrowing the performance advantage over DGEMM. Overall, the results support
selective use of low-precision CPU matrix engines for high-precision GEMM
within the supported input domain, with benefits depending on matrix order,
thread configuration, and the required accuracy.

\appendix
\titleformat{\section}{\large\bfseries}{Appendix~\thesection.}{0.6em}{}
\setcounter{table}{0}
\renewcommand{\thetable}{\thesection.\arabic{table}}
\renewcommand{\theHtable}{appendix.\thesection.\arabic{table}}
\section{Supplementary Numerical Accuracy Results}
\label{app:supplementary-accuracy}

This appendix complements Tables~\ref{tab:accuracy} and~\ref{tab:accuracy-fp64}
with three additional input distributions at $N=1024$, $2048$, $4096$,
$8192$, $16384$, and $32768$. Positive-uniform entries are sampled from
$U(0,1)$. Signed log-uniform entries have independent random signs and base-two
magnitude exponents sampled uniformly from $[-20,20]$, giving magnitudes in
$[2^{-20},2^{20}]$. Cancellation-sensitive inputs pair adjacent
inner-dimension terms so that their leading products cancel, leaving a
contribution approximately $2^{-12}$ of each pair's absolute-product sum.
The component-product selections follow Section~\ref{sec:algorithm}, while
input precisions and reference procedures follow Section~\ref{sec:inputs-metrics}.
These tests characterize distribution sensitivity
rather than establish distribution-independent error bounds.

\subsection{FP32 Accuracy Against an FP64 Reference}

Table~\ref{tab:appendix-fp32} compares both FP32 methods against the common
FP64 reference. For positive-uniform inputs, $e_F$ is
$5.30\times10^{-8}$--$1.49\times10^{-7}$ for AMX-FP32 and
$9.23\times10^{-8}$--$1.69\times10^{-7}$ for SGEMM. AMX-FP32 has lower errors
in both metrics for $N\le4096$, while SGEMM has lower $e_F$ for $N\ge8192$
and lower $e_{\mathrm{scaled}}$ at $N=32768$. Here the exact product equals
$|A||B|$, so cancellation does not amplify relative errors.
Consequently, $e_F$ describes aggregate relative accuracy and
$e_{\mathrm{scaled}}$ approximates the largest entrywise relative error.
The growth of both AMX-FP32 metrics with $N$ is consistent with the greater
amount of FP32 accumulation required for larger products; BF16 decomposition recovers input information
but does not eliminate accumulation rounding. The crossover shows comparable
accuracy rather than a uniform advantage for either method.

For signed log-uniform inputs, AMX-FP32 gives
$e_F=(1.06$--$2.36)\times10^{-7}$, lower than SGEMM at every tested order,
while its scaled errors do not exceed $3.44\times10^{-7}$. The latter slightly
exceed SGEMM's at the two largest orders. Lower Frobenius-norm error does not
necessarily imply a lower maximum scaled error. The six-product method retains accuracy comparable to
SGEMM across this wider input range, although the scaled errors are larger
than for $U(-1,1)$.

For cancellation-sensitive inputs, AMX-FP32 gives
$e_F=(1.43$--$1.48)\times10^{-5}$, compared with
$(6.63$--$6.85)\times10^{-5}$ for SGEMM. With a common reference and
normalization, AMX-FP32's Frobenius error is approximately $1/4.6$ of SGEMM's.
The larger $e_F$ of both methods reflects the deliberately small reference
product relative to $|A||B|$: rounding and truncation errors are magnified
when measured relative to the residual after cancellation. AMX-FP32's scaled
error decreases from $6.04\times10^{-10}$ to $1.40\times10^{-10}$ while $e_F$
remains nearly constant, illustrating that a smaller scaled error does not
by itself imply improved relative accuracy of the cancelled result.
Thus, AMX-FP32 is more accurate than SGEMM in this stress test, but neither
method maintains the $10^{-7}$ normwise error level seen for the other inputs.

\begin{table}[!htbp]
  \centering
  \caption{FP32 numerical accuracy for three additional input distributions.
  AMX-FP32 and SGEMM use identical FP32 inputs and the same FP64 DGEMM reference
  computed after exact widening of those inputs.}
  \label{tab:appendix-fp32}
  \small
  \setlength{\tabcolsep}{5.0pt}
  \renewcommand{\arraystretch}{1.15}
  \begin{tabular}{@{}c cc cc@{}}
    \toprule
    \multirow{2}{*}{$N$}
    & \multicolumn{2}{c}{AMX-FP32} & \multicolumn{2}{c}{oneMKL SGEMM} \\
    \cmidrule(lr){2-3}\cmidrule(lr){4-5}
    & $e_F$ & $e_{\mathrm{scaled}}$ & $e_F$ & $e_{\mathrm{scaled}}$ \\
    \midrule
    \multicolumn{5}{c}{Positive uniform: $U(0,1)$} \\
    \midrule
    1024 & $5.30\times10^{-8}$ & $2.48\times10^{-7}$ & $1.69\times10^{-7}$ & $8.38\times10^{-7}$ \\
    2048 & $5.76\times10^{-8}$ & $2.86\times10^{-7}$ & $1.31\times10^{-7}$ & $6.61\times10^{-7}$ \\
    4096 & $6.32\times10^{-8}$ & $3.10\times10^{-7}$ & $1.02\times10^{-7}$ & $5.43\times10^{-7}$ \\
    8192 & $9.77\times10^{-8}$ & $4.07\times10^{-7}$ & $9.23\times10^{-8}$ & $4.98\times10^{-7}$ \\
    16384 & $1.08\times10^{-7}$ & $5.89\times10^{-7}$ & $9.97\times10^{-8}$ & $7.57\times10^{-7}$ \\
    32768 & $1.49\times10^{-7}$ & $7.59\times10^{-7}$ & $1.28\times10^{-7}$ & $6.93\times10^{-7}$ \\
    \midrule
    \multicolumn{5}{c}{Signed log-uniform} \\
    \midrule
    1024 & $1.06\times10^{-7}$ & $3.21\times10^{-7}$ & $1.69\times10^{-7}$ & $6.82\times10^{-7}$ \\
    2048 & $1.30\times10^{-7}$ & $3.44\times10^{-7}$ & $1.76\times10^{-7}$ & $4.72\times10^{-7}$ \\
    4096 & $1.40\times10^{-7}$ & $2.67\times10^{-7}$ & $1.83\times10^{-7}$ & $3.68\times10^{-7}$ \\
    8192 & $1.58\times10^{-7}$ & $2.69\times10^{-7}$ & $1.93\times10^{-7}$ & $2.80\times10^{-7}$ \\
    16384 & $1.88\times10^{-7}$ & $2.84\times10^{-7}$ & $2.11\times10^{-7}$ & $2.34\times10^{-7}$ \\
    32768 & $2.36\times10^{-7}$ & $2.72\times10^{-7}$ & $2.42\times10^{-7}$ & $2.55\times10^{-7}$ \\
    \midrule
    \multicolumn{5}{c}{Cancellation-sensitive} \\
    \midrule
    1024 & $1.45\times10^{-5}$ & $6.04\times10^{-10}$ & $6.70\times10^{-5}$ & $2.89\times10^{-9}$ \\
    2048 & $1.48\times10^{-5}$ & $4.49\times10^{-10}$ & $6.85\times10^{-5}$ & $2.01\times10^{-9}$ \\
    4096 & $1.43\times10^{-5}$ & $3.28\times10^{-10}$ & $6.63\times10^{-5}$ & $1.65\times10^{-9}$ \\
    8192 & $1.45\times10^{-5}$ & $2.42\times10^{-10}$ & $6.70\times10^{-5}$ & $1.15\times10^{-9}$ \\
    16384 & $1.44\times10^{-5}$ & $1.98\times10^{-10}$ & $6.67\times10^{-5}$ & $8.86\times10^{-10}$ \\
    32768 & $1.44\times10^{-5}$ & $1.40\times10^{-10}$ & $6.67\times10^{-5}$ & $6.11\times10^{-10}$ \\
    \bottomrule
  \end{tabular}
\end{table}
\FloatBarrier

\subsection{FP64 Agreement with DGEMM}

Table~\ref{tab:appendix-fp64} retains DGEMM as the comparison baseline for
FP64 inputs. For positive-uniform inputs, AMX-FP64-6 has $e_F$ on the order
of $10^{-10}$, while AMX-FP64-10 reaches
$2.39\times10^{-13}$--$1.36\times10^{-12}$. Their decreasing $e_F$ with $N$
is consistent with partial cancellation of signed error contributions within
each dot product, while the positive reference terms accumulate without
cancellation. The 15- and 21-product variants
reach discrepancies of approximately $10^{-15}$ and $10^{-16}$--$10^{-15}$,
respectively, with diminishing improvement from additional products.

For signed log-uniform inputs, the 6-, 10-, 15-, and 21-product variants
give $e_F$ of approximately $8.3\times10^{-8}$, $2.4\times10^{-10}$,
$6.2\times10^{-13}$, and $(1.6$--$2.1)\times10^{-15}$, respectively.
AMX-FP64-6 has a scaled discrepancy of $1.09\times10^{-6}$ at $N=1024$,
whereas the 21-product variant keeps it at or below $1.54\times10^{-14}$.
Small normwise discrepancies thus need not imply equally small local ones.

For cancellation-sensitive inputs, increasing the product count reduces $e_F$
relative to DGEMM from approximately $5.6\times10^{-5}$ to
$1.3\times10^{-7}$, $2.8\times10^{-10}$, and $6.4\times10^{-13}$.
Both metrics improve with product count, while the larger normwise
discrepancies reflect the small reference product relative to $|A||B|$.
Across these cases, retaining more products improves agreement with DGEMM;
because DGEMM is itself a rounded computation, these comparisons do not
separate the two implementations' errors relative to the exact product.

\begin{table}[!htbp]
  \centering
  \caption{AMX-FP64 discrepancies relative to oneMKL DGEMM for three additional
  input distributions. All variants and DGEMM use identical FP64 inputs within
  each case; these references are distinct from the widened FP32-input references
  used in Table~\ref{tab:appendix-fp32}.
  Entries use scientific e notation: \textnormal{1.49e-08} denotes
  $1.49\times10^{-8}$.}
  \label{tab:appendix-fp64}
  \footnotesize
  \setlength{\tabcolsep}{3.5pt}
  \renewcommand{\arraystretch}{1.15}
  \begin{tabular}{@{}r rr rr rr rr@{}}
    \toprule
    \multirow{2}{*}{$N$}
    & \multicolumn{2}{c}{AMX-FP64-6} & \multicolumn{2}{c}{AMX-FP64-10}
    & \multicolumn{2}{c}{AMX-FP64-15} & \multicolumn{2}{c}{AMX-FP64-21} \\
    \cmidrule(lr){2-3}\cmidrule(lr){4-5}\cmidrule(lr){6-7}\cmidrule(lr){8-9}
    & $e_F$ & $e_{\mathrm{scaled}}$ & $e_F$ & $e_{\mathrm{scaled}}$
    & $e_F$ & $e_{\mathrm{scaled}}$ & $e_F$ & $e_{\mathrm{scaled}}$ \\
    \midrule
    \multicolumn{9}{c}{Positive uniform: $U(0,1)$} \\
    \midrule
    1024 & \textnormal{6.24e-10} & \textnormal{3.16e-09} & \textnormal{1.36e-12} & \textnormal{6.64e-12} & \textnormal{2.93e-15} & \textnormal{1.48e-14} & \textnormal{3.90e-16} & \textnormal{2.20e-15} \\
    2048 & \textnormal{4.39e-10} & \textnormal{2.28e-09} & \textnormal{9.65e-13} & \textnormal{5.48e-12} & \textnormal{2.08e-15} & \textnormal{1.18e-14} & \textnormal{3.72e-16} & \textnormal{2.19e-15} \\
    4096 & \textnormal{3.10e-10} & \textnormal{1.78e-09} & \textnormal{6.81e-13} & \textnormal{3.81e-12} & \textnormal{1.49e-15} & \textnormal{7.85e-15} & \textnormal{4.23e-16} & \textnormal{2.41e-15} \\
    8192 & \textnormal{2.19e-10} & \textnormal{1.24e-09} & \textnormal{4.79e-13} & \textnormal{2.67e-12} & \textnormal{1.14e-15} & \textnormal{6.44e-15} & \textnormal{6.08e-16} & \textnormal{3.74e-15} \\
    16384 & \textnormal{1.55e-10} & \textnormal{9.07e-10} & \textnormal{3.39e-13} & \textnormal{2.07e-12} & \textnormal{9.73e-16} & \textnormal{5.59e-15} & \textnormal{7.89e-16} & \textnormal{4.59e-15} \\
    32768 & \textnormal{1.09e-10} & \textnormal{6.55e-10} & \textnormal{2.39e-13} & \textnormal{1.44e-12} & \textnormal{1.02e-15} & \textnormal{6.21e-15} & \textnormal{9.76e-16} & \textnormal{6.21e-15} \\
    \midrule
    \multicolumn{9}{c}{Signed log-uniform} \\
    \midrule
    1024 & \textnormal{8.37e-08} & \textnormal{1.09e-06} & \textnormal{2.38e-10} & \textnormal{3.34e-09} & \textnormal{6.23e-13} & \textnormal{1.06e-11} & \textnormal{1.64e-15} & \textnormal{1.54e-14} \\
    2048 & \textnormal{8.34e-08} & \textnormal{3.16e-07} & \textnormal{2.37e-10} & \textnormal{1.25e-09} & \textnormal{6.21e-13} & \textnormal{3.22e-12} & \textnormal{1.65e-15} & \textnormal{8.95e-15} \\
    4096 & \textnormal{8.33e-08} & \textnormal{1.82e-07} & \textnormal{2.37e-10} & \textnormal{6.12e-10} & \textnormal{6.20e-13} & \textnormal{1.22e-12} & \textnormal{1.68e-15} & \textnormal{3.78e-15} \\
    8192 & \textnormal{8.33e-08} & \textnormal{1.03e-07} & \textnormal{2.37e-10} & \textnormal{2.92e-10} & \textnormal{6.20e-13} & \textnormal{7.92e-13} & \textnormal{1.75e-15} & \textnormal{2.05e-15} \\
    16384 & \textnormal{8.33e-08} & \textnormal{6.48e-08} & \textnormal{2.37e-10} & \textnormal{1.79e-10} & \textnormal{6.20e-13} & \textnormal{5.09e-13} & \textnormal{1.88e-15} & \textnormal{1.70e-15} \\
    32768 & \textnormal{8.33e-08} & \textnormal{4.19e-08} & \textnormal{2.37e-10} & \textnormal{1.21e-10} & \textnormal{6.20e-13} & \textnormal{3.27e-13} & \textnormal{2.11e-15} & \textnormal{1.84e-15} \\
    \midrule
    \multicolumn{9}{c}{Cancellation-sensitive} \\
    \midrule
    1024 & \textnormal{5.64e-05} & \textnormal{2.32e-09} & \textnormal{1.27e-07} & \textnormal{5.14e-12} & \textnormal{2.76e-10} & \textnormal{1.26e-14} & \textnormal{6.47e-13} & \textnormal{2.35e-17} \\
    2048 & \textnormal{5.65e-05} & \textnormal{1.69e-09} & \textnormal{1.31e-07} & \textnormal{4.16e-12} & \textnormal{2.83e-10} & \textnormal{9.75e-15} & \textnormal{6.61e-13} & \textnormal{2.08e-17} \\
    4096 & \textnormal{5.53e-05} & \textnormal{1.29e-09} & \textnormal{1.25e-07} & \textnormal{2.92e-12} & \textnormal{2.73e-10} & \textnormal{6.30e-15} & \textnormal{6.38e-13} & \textnormal{1.59e-17} \\
    8192 & \textnormal{5.57e-05} & \textnormal{1.05e-09} & \textnormal{1.26e-07} & \textnormal{2.15e-12} & \textnormal{2.76e-10} & \textnormal{4.74e-15} & \textnormal{6.42e-13} & \textnormal{1.15e-17} \\
    16384 & \textnormal{5.52e-05} & \textnormal{6.80e-10} & \textnormal{1.26e-07} & \textnormal{1.64e-12} & \textnormal{2.74e-10} & \textnormal{3.32e-15} & \textnormal{6.42e-13} & \textnormal{8.13e-18} \\
    32768 & \textnormal{5.53e-05} & \textnormal{5.16e-10} & \textnormal{1.25e-07} & \textnormal{1.19e-12} & \textnormal{2.74e-10} & \textnormal{2.68e-15} & \textnormal{6.41e-13} & \textnormal{6.02e-18} \\
    \bottomrule
  \end{tabular}
\end{table}
\FloatBarrier

\bibliographystyle{unsrtnat}
\bibliography{references}

\end{document}